\documentclass[]{jfm}

\usepackage{graphicx}
\usepackage{newtxtext}
\usepackage{newtxmath}
\usepackage{natbib}
\usepackage{hyperref}
\hypersetup{
    colorlinks = true,
    urlcolor   = blue,
    citecolor  = black,
}

\newcommand{\RomanNumeralCaps}[1]
\linenumbers

\title{Hindered settling of log-normally distributed particulate suspensions: theoretical models vs.  Stokesian  simulations }

\author{Heng Li\aff{1}
  \and Lorenzo Botto\aff{1}
  \corresp{\email{l.botto@tudelft.nl}}
 }

\affiliation{\aff{1}Process and Energy Department, ME Faculty of Mechanical Engineering, TU Delft, 2628 CD Delft, The Netherlands
}

\begin{document}
\maketitle
%%%%%%%%%%%%%%%%%%%%%%%%%%%%%%%%%%%%%%%%%%%%%
\begin{abstract}
Settling velocity statistics for dilute, non-Brownian suspensions of polydisperse spheres having a log-normal size distribution are analysed by Stokesian Dynamics, as a function of the total volume fraction and width of the size distribution. Several hundred instantaneous configurations are averaged to obtain reliable statistics. Average velocities for each particle class are compared to the models proposed by Batchelor, Richardson $\&$ Zaki, Davis $\&$ Gecol, and Masliyah-Lockett-Bassoon (MLB). Batchelor's model is shown to give reasonably accurate predictions when the volume fraction is within 5\%. Because of its complexity, this  model is however hardly used in practice, so lower-order models are needed. We found that while the other hindered settling models can give reasonably accurate predictions of the velocity of the largest particles,  all of them overestimate - in certain cases by a large margin - the velocity of the smaller particles.  By computing the fluid-particle velocity slip for each particle class and using Batchelor's model, we explain why predicting the lower tail of the particle size distribution is challenging, and propose possible avenues for model improvement. The analysis of velocity fluctuations suggest quantitative similarities between velocity fluctuations in monodisperse and polydisperse suspensions.

%The finite-Reynolds-number three-dimensional flow in a channel bounded by one and
%two parallel porous walls is studied numerically. The porous medium is modelled by
%spheres in a simple cubic arrangement. Detailed results on the flow structure and
%the hydrodynamic forces and couple acting on the sphere layer bounding the porous
%medium are reported and their dependence on the Reynolds number illustrated. It
%is shown that, at finite Reynolds numbers, a lift force acts on the spheres, which
%may be expected to contribute to the mobilization of bottom sediments. The results
%for the slip velocity at the surface of the porous layers are compared with the
%phenomenological Beavers–Joseph model. It is found that the values of the slip
%coefficient for pressure-driven and shear-driven flow are somewhat different, and also
%depend on the Reynolds number. A modification of the relation is suggested to deal
%with these features. The Appendix provides an alternative derivation of this modified
%relation

\end{abstract}

\begin{keywords}
Authors should not enter keywords on the manuscript, as these must be chosen by the author during the online submission process and will then be added during the typesetting process (see \href{https://www.cambridge.org/core/journals/journal-of-fluid-mechanics/information/list-of-keywords}{Keyword PDF} for the full list).  Other classifications will be added at the same time.
\end{keywords}

%%%%%%%%%%%%%%%%%%%%%%%%%%%%%%%%%%%%%%%%%%%%
\section{Introduction}
\label{sec:intro}
The prediction of the settling velocity of polydisperse suspensions is crucially important in applications, such as wastewater treatment \citep{he2021innovative}, food processing \citep{wang2020improving}, nanoparticle sorting \citep{bonaccorso2013sorting}, particle size characterisation \citep{papuga2021we}, materials recycling \citep{wolf2021centrifugation}, and sediment transport modelling \citep{dorrell2010sedimentation}. Despite decades of research on polydispersed suspension this field still offers interesting scientific problems. A central challenge is the prediction of the settling velocity of each particle class in a polydisperse system. This information is crucially important, not only because the velocity of each class dictates the particle concentration profile, but also because only by knowing the velocity of each class it is possible to separate particles by size. The accurate prediction of the class-averaged particle velocity  has recently become important because of the need for accurate size fractionation of micro and nanoparticles \citep{bonaccorso2013sorting,backes2016production}. Furthermore, from the knowledge of the class-averaged particle velocity  one can measure the particle size distribution from the time evolution of the concentration profile \citep{papuga2021we}, which is the principle underlying the functioning of an analytical centrifuge \citep{chaturvedi2018measuring}. The current work aims to analyse the validity of several models used for the prediction of the class-averaged velocity, comparing against simulation results.

For a Stokesian suspension of polydisperse spheres grouped into $m$ distinct particle classes, the average settling velocity of the $i$-th class can be written as $\langle u_i \rangle = u_{St,i} h_i(\boldsymbol{\phi})$, where $u_{St,i}=\frac{2}{9\mu} a_i^2 (\rho_p-\rho_f)g$ is the single-particle Stokes velocity of the $i$-th class, $h_i(\boldsymbol{\phi})$ is the hindered settling function of that class, and $\boldsymbol{\phi}=(\phi_1,\phi_2,...,\phi_m)$ is the vector of  volume fractions \citep{davis1985sedimentation}; $a_i$ is the particle radius, $\mu$ is the fluid viscosity, and $\rho_p - \rho_f$ is the density difference between the particles and the fluid. The literature reports  several models for $h_i(\boldsymbol{\phi})$, as reviewed by \citet{berres2005applications}.

\citet{batchelor1982sedimentation}  showed that in the dilute limit the hindered settling function can be written as
\begin{equation}
    h_i(\boldsymbol{\phi})=1+\sum_{j=1}^{m} S_{ij}\phi_j,
    \label{batchelor}
\end{equation}
and using the pair-wise interaction approximation developed analytical solution for
the sedimentation coefficients $S_{ij}$ for different size and density ratios \citep{batchelor1982sedimentationb}. In equation (\ref{batchelor}) $S_{ii}=-6.55$, so  the hindered settling function for monodispersed suspensions is recovered for $m=1$ \citep{batchelor1972sedimentation}.

\citet{davis1994hindered} proposed the following semi-empirical extension of Batchelor's formula:
\begin{equation}
    h_i(\boldsymbol{\phi})= (1-\phi)^{-S_{ii}} \left(1+\sum_{j=1}^{m} (S_{ij}-S_{ii})\phi_j \right),
    \label{dg}
\end{equation}
where $\phi=\sum \phi_j$  is the total volume fraction, and the coefficients $S_{ij}$ are defined as in equation (\ref{batchelor}). for For $\phi \rightarrow 0 $, equation (\ref{dg})  recovers (\ref{batchelor}).  For $m=1$,  equation (\ref{dg}) reduces to a power-law form, with exponent  $-S_{ii}$, similar to the expression of \citet{richardson1954sedimentation}.  

The models of Batchelor and Davis $\&$ Gecol have been tested in experiments and simulations of bidisperse suspensions \citep{davis1988hindered,al1992sedimentation,abbas2006dynamics,wang2015short,chen2023characterising}. However, these models are rarely used in practice because they contain a large number of coefficients. Batchelor's model furthermore does not smoothly converge to a form that can handle large solid concentrations, and therefore cannot be used in one-dimensional simulation where the concentration increases from dilute in the supernatant to concentrated in the region immediately above the sediment-supernatant interface.  Simpler expressions have therefore been developed for practical predictions of the hindered settling of polydisperse suspensions.

Some authors \citep{davis1988spreading,abeynaike2012experimental,vowinckel2019settling,chen2023characterising} have adapted the model of \citet{richardson1954sedimentation} to the velocity of each class by writing
\begin{equation}
h_i(\boldsymbol{\phi}) = (1-\phi)^n.
\label{r-z}
\end{equation}
This model predicts different velocities for different particle radii $a_i$, because $h_i$ contains the single-particle Stokes formula at denominator. In the current paper, this hindered settling formula will be referred to as Richardson-Zaki's model for polydispersed suspensions. The exponent $n$ is obtained from an empirical fit. For monodispersed suspensions, \citet{richardson1954sedimentation} originally proposed $n \approx 5$, but the exact value is still a subject of debate. For example, \citet{brzinski2018observation} demonstrated that the value of $n$ depends on the particle Peclet number, and there are two branches in the sedimentation curve which are best fitted by $n\approx5.6$ for monodisperse Brownian spheres, and  $n\approx4.48$ for monodisperse non-Brownian spheres.

 \citet{masliyah1979hindered} and \citet{lockett1979sedimentation} proposed the following hindered settling formula:
\begin{equation}
    h_i(\boldsymbol{\phi})= (1-\phi)^{n-1} \left(1-\sum_{j=1}^{m} \left(\frac{a_j}{a_i}\right)^2\phi_j \right),
    \label{mlb}
\end{equation}
where $a_j/a_i$ is the ratio between the radii of the $j$-th and the $i$-th species. The function $\left(1-\sum_{j=1}^{m} \left(\frac{a_j}{a_i}\right)^2\phi_j \right)$ is the hindered settling function obtained by including the effect of volume fraction on the fluid-solid slip velocity, and neglecting the effect of hydrodynamic interactions on the drag force experienced by each particle. The prefactor $(1-\phi)^{n-1}$ estimates the effect of hydrodynamic interactions on the drag force (see Appendix \ref{appA}, where equation (\ref{mlb}) is re-derived).

The Masliyah-Lockett-Bassoon (MLB) model is favoured in engineering applications \citep{xue2003modeling,berres2005applications,dorrell2010sedimentation}. It is easy to tune, since it has only one fitting parameter. Using the MLB model for the stability analysis of settling size-polydispersed and equal-density spheres gives no unphysical lateral segregation and streamers, which are instead obtained using Davis \& Gecol's model \citep{buerger2002model,tory2003strongly,berres2005applications}. The MLB model has been  validated by comparison of predicted and measured concentration profiles \citep{xue2003modeling,berres2005applications}. However, this validation is not complete, because the particle concentration is a convolution of the velocities of the different size classes. Therefore a reasonably accurate prediction of concentration does not necessarily imply that the velocity of each class has been predicted correctly. Direct validation of the settling rates predicted by the MLB model for all size classes has not been published.

Despite large recent interest in the modelling of suspensions of wide and continuous size distributions \citep{pednekar2018bidisperse,gonzalez2021impact,howard2022bidisperse,malbranche2023shear,lavrenteva2024shear}, there is very limited data on the settling of polydisperse suspensions with many size classes. Most physical experiments have been carried out for bidisperse or tridisperse suspensions \citep{lockett1979sedimentation,davis1988hindered,al1992sedimentation,davis1994hindered,chen2023characterising}. In these experiments, the largest size ratio between the two species was around 4, and the {velocity} of the largest size class {was} only measured in the homogeneous region where a mixture of all size classes was present. Numerical simulations have been carried out for bidisperse suspensions using Stokesian dynamics \citep{revay1992numerical,cunha2002modeling,wang2015short} and the force coupling method \citep{abbas2006dynamics}, with size ratio up to 4. Simulations of sedimentation of suspensions with a log-normal distribution have been carried out by \citet{vowinckel2019settling} {in a domain bounded by top and bottom walls. Their objective was to study the effect of cohesive force on the transient settling process. The velocity of each size class was not quantified.    

In this paper, we analyse settling velocity statistics for polydisperse suspensions of non-Brownian spheres using Stokesian dynamics simulations. The size distribution is log-normal. Of all the particle size distributions, log-normal distributions are the most interesting because of their ubiquitous presence in applications \citep{vowinckel2019settling,di2022influence,rettinger2022rheology}. We vary the volume fraction and ratio between the standard deviation and mean value of the particle size distribution. All the particles have the same mass density. The volume fraction ranges from 0.01 to 0.1. The largest size ratio between two classes is 5 and up to 9 classes are considered. To reduce the large statistical error, following other authors \citep{revay1992numerical,cunha2002modeling,abbas2006dynamics,wang2015short}, we produce converged particle velocity statistics by generating a large number of random fixed particle configurations inside a triply periodic box and ensemble-averaging over all the configurations.  

%We analyse the probability distributions, average values and fluctuations of the particle velocities. We compare the hindered settling functions of each size class calculated from our simulations with the predictions from different models to verify their accuracies. Of the three aforementioned models, we primarily focus more on analysing the MLB model due to its simplicity and broad applications. Accurate hindered settling function models are needed not only to describe the sedimentation process, but also to model polydisperse suspension flows in general as closures for the suspension balance model \citep{nott2011suspension,howard2022bidisperse}.

%%%%%%%%%%%%%%%%%%%%%%%%%%%%%%%%%%%%%%%%%%%%
\section{Numerical approach and validation}
\label{sec:approach}

\begin{figure}
  \centerline{\includegraphics[width=0.6\textwidth]{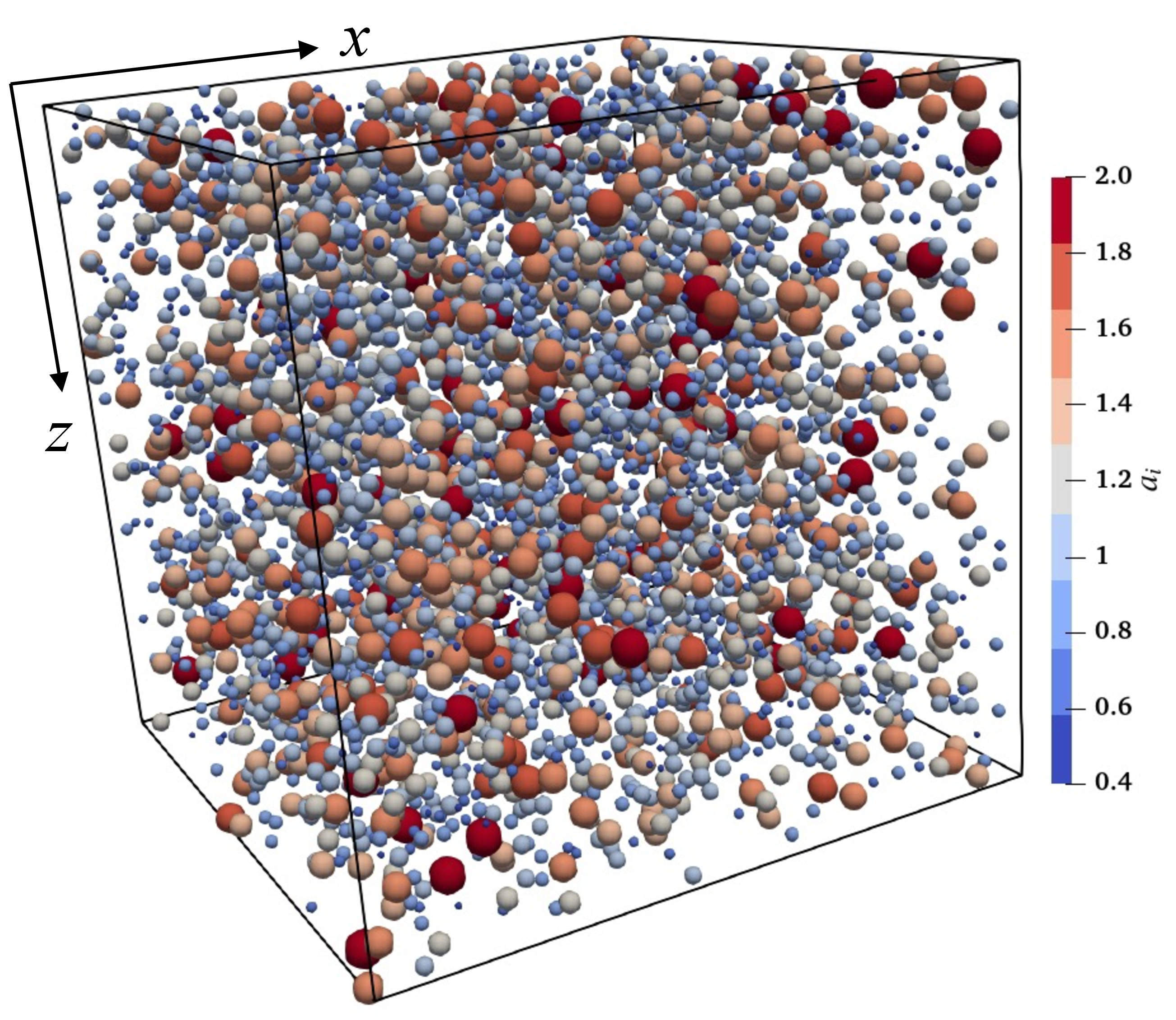}}% Images in 100% size
  \caption{A configuration for volume fraction $\phi = 0.05$ and polydispersity parameter $\alpha = 0.4$. The spheres are coloured according to their radii.}
\label{fig:1}
\end{figure}

Consider a polydisperse suspension of $N$ spheres having the same density but different radii. The $N$ spheres are divided into  $m$ size classes. The radius of size class $i$ is $a_i$. Each sphere in class $i$ is subjected to a  force $\boldsymbol{F}_i=\frac{4}{3} \pi a_i^3 (\rho_p-\rho_f) \boldsymbol{g}$, which includes the particle weight and buoyancy; $\rho_p$ and $\rho_f$ are the densities of the spheres and the fluid, respectively, and $\boldsymbol{g}$ is the gravitational acceleration. The single-particle Stokes velocity corresponding to each class is $\boldsymbol{u}_{St,i}=\frac{2}{9\mu} a_i^2 (\rho_p-\rho_f)\boldsymbol{g}$, where $\mu$ is the dynamic viscosity of the fluid. In the current work, the particle velocity statistics are calculated from each configuration of the spheres by first averaging over the particles in the computational domain and then ensemble-averaging over statistically identical configurations. Each configuration is generated by randomly placing the spheres one by one inside the computational domain, ensuring that each placement gives no overlap between the spheres   \citep{revay1992numerical,wang2015short,cheng2023hydrodynamic}. One such configuration is shown in figure \ref{fig:1}. In our coordinate system, gravity is aligned in the $z$ direction, also referred to as vertical direction in the following.  The horizontal direction corresponds to the $x$ and $y$ coordinates.  

To calculate the velocities of individual particles, a basic version of the Stokesian Dynamics method is adopted \citep{brady1988dynamic,brady1988stokesian}. The velocities of the spheres are calculated by solving the mobility problem 
\begin{equation}
    \boldsymbol{U}-\langle \boldsymbol{u} \rangle = \mathsfbi{M} \boldsymbol{F},
    \label{mobility}
\end{equation}
where $\boldsymbol{U}$ is the $3N$ vector containing the velocities of the spheres, $\boldsymbol{F}$ is the $3N$ vector containing the gravitational forces acting on the spheres (these forces include the particle weight and the  buoyancy force), and $\mathsfbi{M}$ is the $3N\times3N$ mobility matrix \citep{kim2013microhydrodynamics}. In equation (\ref{mobility}), $\langle \boldsymbol{u} \rangle$ is the average translational velocity of the suspension. In our simulations $\langle \boldsymbol{u} \rangle = \boldsymbol{0}$ because of the zero volume flux condition of batch sedimentation \citep{berres2005applications}. Note that in the current work only velocity-force coupling is considered, i.e. the stresslet and other force moments are not considered. \citet{brady1988sedimentation} showed that in a  sedimenting suspension the inclusion of the stresslet changes the settling rate negligibly. Because we work in the relatively dilute limit, short-range lubrication are also neglected.

The mobility matrix $\mathsfbi{M}$ depends on the positions and radii of the spheres. We used the Rotne-Prager approximation for this term \citep{rotne1969variational,zuk2014rotne}. This approximation has been shown to give accurate predictions of the sedimentation velocities of suspensions from dilute to relatively dense \citep{brady1988sedimentation}.  Triply periodic boundary conditions are applied to the simulation box. The mobility matrix is constructed using the Ewald summation technique by splitting the mobility matrix into a real-space part and a wave-space part \citep{beenakker1986ewald}. Explicit formulae for the mobility matrix for a polydisperse suspension can be found in \citet{beenakker1986ewald} and \citet{hase2001calculation}. As characteristic length and velocity scales, we choose the mean  particle radius  $\langle a \rangle$ and the single particle Stokes velocity corresponding to $\langle a \rangle$. To make forces non-dimensional we use the effective weight of the mean particle, $\frac{4}{3} \pi \langle a \rangle^3 (\rho_p-\rho_f) g$.

\begin{figure}
  \centerline{\includegraphics{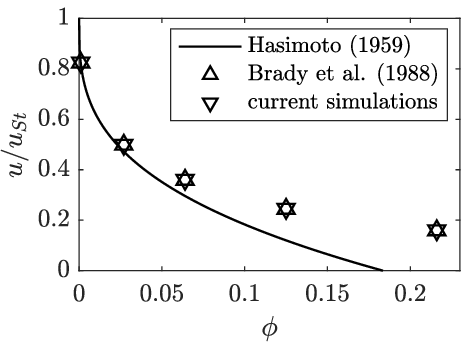}}% Images in 100% size
  \caption{Normalized settling velocity vs. volume fraction for a simple cubic array of monodisperse spheres. The line is the point-force solution of \citet{hasimoto1959periodic}. Upward triangles are the numerical results of \citet{brady1988dynamic}.}
\label{fig:2}
\end{figure}

In figure \ref{fig:2}  numerical predictions for a single sphere in a triply periodic cubic box are plotted against Hasimoto’s analytical solution \citep{hasimoto1959periodic} and the simulation results of \citet{brady1988dynamic}. The volume fraction of the simple cubic array is varied by varying the size of the box. Based on the point-force assumption, \citet{hasimoto1959periodic} derived $u/u_{St}=1-1.7601\phi^{1/3}$ for $\phi \ll 1$, where $u_{St}$ is the Stokes velocity of the sphere. \citet{brady1988dynamic} used Stokesian Dynamics with different approximations for the mobility matrix. The results of \citet{brady1988dynamic} shown in figure \ref{fig:2} correspond to simulations in the Rotne-Prager approximation.  As seen from figure \ref{fig:2}, our results match exactly those of \citet{brady1988dynamic} and converge to Hasimoto’s solution for $\phi \rightarrow 0$. This test validates our implementation of the Ewald summation for the periodic boundary conditions.

\begin{figure}
  \centerline{\includegraphics[width=1.0\textwidth]{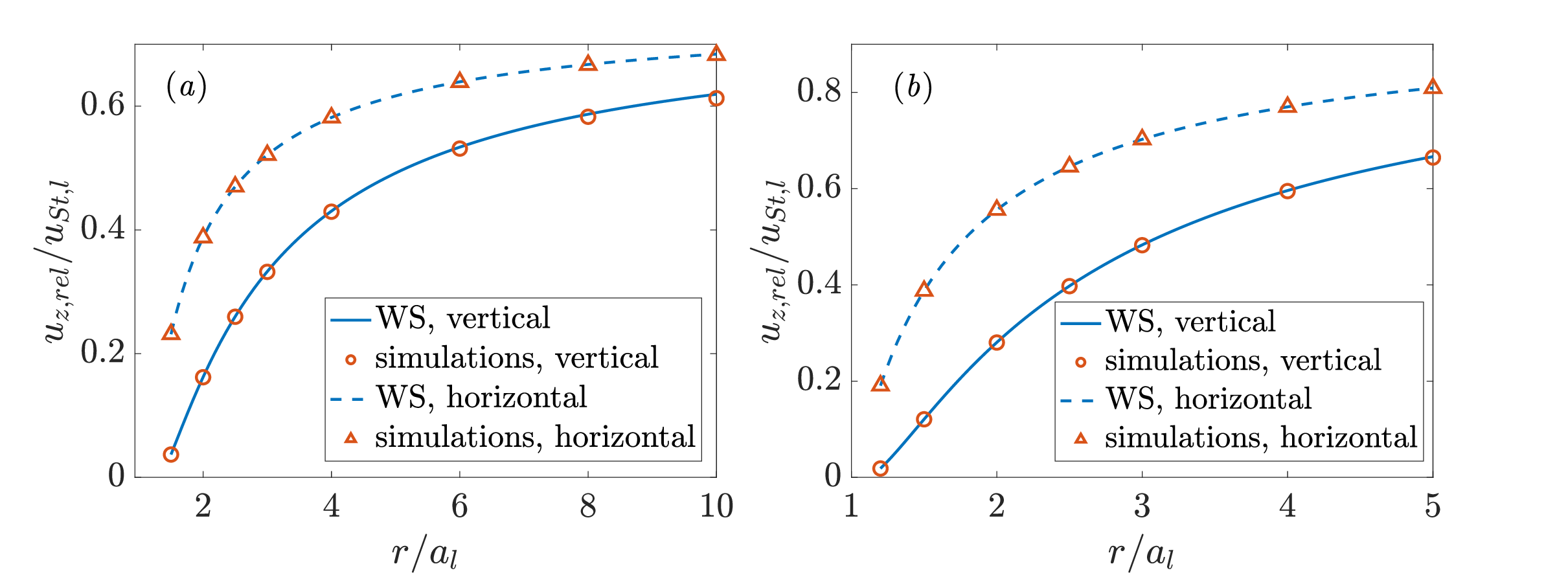}}% Images in 100% size
  \caption{Normalized relative settling velocity for a pair of spheres as a function of the centre-to-centre distance for (\textit{a}) size ratio 2\protect\\ and (\textit{b}) size ratio 5. Results of current simulations are shown as symbols, and analytical results of \citet{wacholder1974hydrodynamic} are shown as lines.}
\label{fig:3}
\end{figure}

In figure \ref{fig:3}, the normalized relative settling velocity is shown as a function of the normalized centre-to-centre distance between two unequal spheres with size ratio 2 and 5, respectively. In our simulations, the radius of the large sphere is fixed to $a_l=2$. The radius of the small sphere is $a_s=1$ and 0.4 for size ratio 2 and 5, respectively (these values are chosen because the largest  radius is 2 and the largest size ratio  is 5 in the polydisperse simulations analysed in this paper). The relative settling velocity between the two spheres is normalized by the Stokes velocity of the large sphere. The centre-to-centre distance is normalized by the radius of the large sphere.  In figure \ref{fig:3}, symbols are results from our simulations, and lines correpond to the  asymptotic solution of \citet{wacholder1974hydrodynamic}, in which only  far-field hydrodynamic interactions were considered. It can be seen that our results match  the analytical solution for both vertically and horizontally aligned pairs. 

The current paper discusses results for bidisperse suspensions and polydisperse suspensions with more than two classes, also comparing with the monodisperse case. For the monodisperse case, the radius of the spheres is $a=1$. For the  bidisperse case, two size ratios are considered: $a_2/a_1 =2$ and 5. The radii of the small size classes are $a_1=0.8$ and 0.4 for these two size ratios, respectively. The volume fraction of the small size class is $\phi_1=\frac{3}{11}\phi$ for size ratio 2, and $\phi_1=\frac{1}{76}\phi$ for size ratio 5. These volume fraction ratios are chosen so that the average radius of the spheres is equal to 1.0 for each system. 

\begin{figure}
  \centerline{\includegraphics{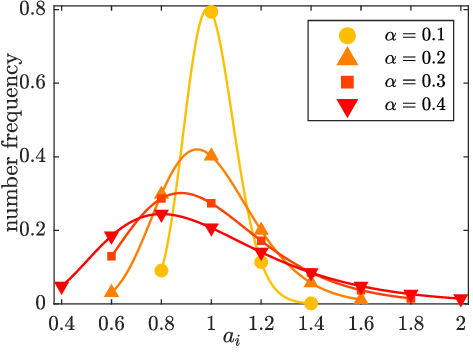}}% Images in 100% size
  \caption{Discrete frequency distributions of particle size for different values of the polydispersity parameter (symbols). Lines indicate the continuous log-normal distributions that fit the discrete frequency histograms.}
\label{fig:size_dis}
\end{figure}

For the simulations with several size classes, the particle size distribution follows $p(a)=\frac{1}{a\sigma \sqrt{2\pi}} e^{-(\ln{a}-\mu)^2/2\sigma^2}$, where the mean value of the size distribution is $\langle a \rangle = e^{\mu+\sigma^2/2}$ and the standard deviation is $\Delta a=\sqrt{\left( e^{\sigma^2}-1 \right) e^{2\mu+\sigma^2}}$. We define the \textit{polydispersity parameter} as $\alpha=\Delta a/\langle a \rangle$. Four size distributions are considered, with $\langle a \rangle =1$ and $\alpha=$0.1,0.2,0.3 or 0.4. Each distribution is cut at the two ends, resulting in a range  $a_{min} \leq a \leq a_{max}$, where $a_{min}$ and $a_{max}$ are chosen so that at least 95\% of the original distribution falls within this range. The largest size ratio between two spheres is 5. Each radius range is discretized into between 4 and 9 size classes, with a difference of 0.2 between the radii of two adjacent size classes. 

The  discrete number frequency distributions are overlaid on the corresponding continuous distributions in figure \ref{fig:size_dis} . {The frequency of size class $i$ is calculated as $\frac{p(a_i)}{\sum_{j=1}^{m}p(a_j)}$.} {For each value of $\alpha$, the volume fraction $\phi$ ranges from 0.01 to 0.10. For each simulated case, corresponding to a combination of $\alpha$ and $ \phi$, a fixed box size $L=80$ is used and 500 random particle configurations are generated. The total number of spheres in each case varies from 925 to 12223.}

The average velocity of class $i$ is calculated by ensemble-averaging over $M$ configurations  as
\begin{equation}
    \langle \overline{{u}}_{\xi,i} \rangle = \frac{\sum_{k=1}^M  \overline{u}_{\xi,i}^k}{M},
    \label{average}
\end{equation}
where $\xi=1,2,3$ correspond to the three Cartesian coordinates,  $\overline{u}_{\xi,i}$ is the intrinsic volume average of the velocity component ${u}_{\xi,i}$ within one configuration, and $\left < \cdot \right >$ is the ensemble-averaging operator. The intrinsic volume average within class $i$ over  configuration $k$ is $ \overline{u}_{\xi,i}^k=\frac{\sum_{l=1}^{N_i}u_{\xi,i,l}^k}{N_i}$, where $N_i$ is the number of particles in class $i$.  The standard deviation of a certain velocity component within one realisation is calculated as   
$\left(u_{\xi,i}^{\prime}\right)^k=\sqrt{\frac{\sum_{l=1}^{N_i}\left(u_{\xi,i,l}^k- \overline u_{\xi,i}^k\right)^2}{N_i-1}}$. Averaging over many realisations gives an improve estimate of the class-averaged standard deviation. In the bulk of the paper we indicate averages by the bracket symbol, distinguishing between volume and ensemble average when necessary.

%%%%%%%%%%%%%%%%%%%%%%%%%%%%%%%%%%%%%%%%%%%%
\subsection{Relation between the mobility formulation and Batchelor's formula}
In this section we show the connection between the mobility formulation, equation (\ref{mobility}), and Batchelor's formula, equation (\ref{batchelor}). For simplicity of notation, let us consider a specific size class. Without loss of generality we consider class 1. According to (\ref{mobility}) the velocity of the $\alpha$-th sphere in the 1-st size class is
\begin{equation}
    \boldsymbol{u}_{\alpha,1}=\sum_{i=1}^{m}\sum_{\beta=1}^{N_i}\mathsfbi{M}_{\alpha 1,\beta i} \boldsymbol{F}_{i},
    \label{one-velocity}
\end{equation}
where $N_i$ is the number of spheres in the $i$-th class, and $\mathsfbi{M}_{\alpha 1,\beta i}$ is the $3\times 3$ mobility matrix representing the hydrodynamic interaction between the $\alpha$-th sphere in the 1-st class and the $\beta$-th sphere in the $i$-th class \citep{kim2013microhydrodynamics}. Because $\mathsfbi{M}_{\alpha 1,\alpha 1}=(6\pi\mu a_{1})^{-1}$, (\ref{one-velocity}) can be written as
\begin{equation}
    \boldsymbol{u}_{\alpha,1}=\boldsymbol{u}_{St,1}+\sum_{\beta\neq \alpha}\mathsfbi{M}_{\alpha 1,\beta 1}\boldsymbol{F}_{1}+\sum_{i\neq1}\sum_{\beta=1}^{N_i}\mathsfbi{M}_{\alpha 1,\beta i} \boldsymbol{F}_{i}.
    \label{new-one-velocity}
\end{equation}
The average velocity of the 1-st class in this configuration is 
\begin{equation}
    \overline{\boldsymbol{u}}_{1}=\boldsymbol{u}_{St,1}+\frac{1}{N_1}\left(\sum_{\alpha}\sum_{\beta\neq \alpha}\mathsfbi{M}_{\alpha 1,\beta 1}\boldsymbol{F}_{1}+\sum_{\alpha}\sum_{i\neq1}\sum_{\beta=1}^{N_i}\mathsfbi{M}_{\alpha 1,\beta i} \boldsymbol{F}_{i}\right),
    \label{average-velocity}
\end{equation}
but because $\boldsymbol{F}_{i}$ is constant within the same size class we can also write
\begin{equation}
    \overline{\boldsymbol{u}}_{1}=\boldsymbol{u}_{St,1}+\mathsfbi{s}_{11}\boldsymbol{F}_{1}+\sum_{i\neq 1}\mathsfbi{s}_{1i}\boldsymbol{F}_{i},
    \label{new-average}
\end{equation}
where $\mathsfbi{s}_{11}$ and $\mathsfbi{s}_{1i}$ describe the intra-class hydrodynamic interactions (within the 1-st class) and the inter-class hydrodynamic interactions (between the 1-st and the $i$-th classes), respectively. These two matrices can be written as $\mathsfbi{s}_{11}=(N_1-1)\overline{\mathsfbi{M}}_{11}$ and $\mathsfbi{s}_{1i}=N_i\overline{\mathsfbi{M}}_{1i}$, where $\overline{\mathsfbi{M}}_{11}$ and $\overline{\mathsfbi{M}}_{1i}$ are the average two-sphere mobility matrices.  Upon ensemble-averaging, the average velocity of the 1-st size class can be written as 
\begin{equation}
    \langle\boldsymbol{\overline{u}}_{1}\rangle=\boldsymbol{u}_{St,1}+\langle\mathsfbi{s}_{11}\rangle\boldsymbol{F}_{1}+\sum_{i\neq 1}\langle\mathsfbi{s}_{1i}\rangle\boldsymbol{F}_{i}.
    \label{ensemble-average}
\end{equation}
The average velocity component in the gravity direction can be written as  
\begin{equation}
    \frac{\langle u_1 \rangle}{u_{St,1}} = 1 + \frac{9\mu\langle s_{11} \rangle}{2a_1^2 n_1}\phi_1 + \sum_{i\neq1}\frac{9\mu\langle s_{1i} \rangle}{2 a_1^2 n_i}\phi_i,
    \label{non-dimensional}
\end{equation}
where the formula for the single-particle Stokes velocity is used and $n_i$ is the number density of class $i$. The scalar $\langle s_{1i} \rangle$ is the component of $\langle\mathsfbi{s}_{1i}\rangle$ for the velocity-force coupling in the gravity direction. 

Extending equation (\ref{non-dimensional}) to a generic class $i$ yields
\begin{equation}
    \frac{\langle u_i \rangle}{u_{St,i}} = 1 + B_{ii}(\boldsymbol{\phi})\phi_i + \sum_{j\neq i}B_{ij}\left(\boldsymbol{\phi},\frac{a_j}{a_i}\right)\phi_j.
    \label{general}
\end{equation}
The dependence of $B_{ii}$ and $B_{ij}$ on the  volume fraction vector $\boldsymbol{\phi}$ comes from the fact that $\langle s_{ij} \rangle$ depends on the pair distribution functions, and the pair distribution functions in turn depend on the volume fraction of each class. The dependence of $B_{ij}$ on $a_j/a_i$ comes from the dependence of the two-sphere mobility matrix on the size ratio.  For $\phi \rightarrow 0$, $B_{ii}$ is a constant and $B_{ij}=S_{ij}$ is only a function of $a_j/a_i$. In this limit, equation (\ref{general}) recovers Batchelor's expression (\ref{batchelor}).

\section{Hindered settling of monodisperse and bidisperse suspensions}
\label{sec:mono and bi}

\begin{figure}
  \centerline{\includegraphics{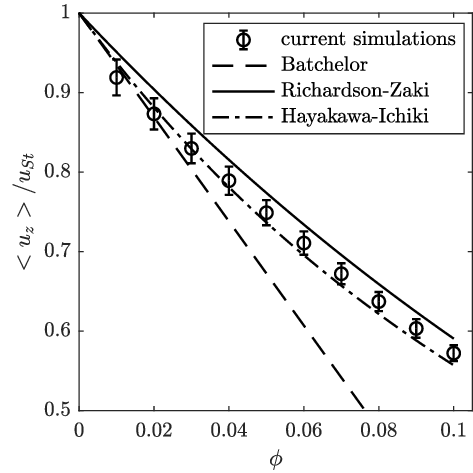}}% Images in 100% size
  \caption{Monodispese case: average settling velocity, normalized by the single-particle Stokes velocity, versus volume fraction}
\label{fig:mono_mean}
\end{figure}

The normalised average settling velocity $\langle u_z \rangle/u_{St}$  for the monodisperse suspension is plotted in figure \ref{fig:mono_mean} as a function of $\phi$. We include in the plot the Richardson-Zaki correlation $(1-\phi)^n$ \citep{richardson1954sedimentation} for $n=5$, the Batchelor model $1+S\phi$ \citep{batchelor1972sedimentation} assuming $S=-6.55$ and the Hayakawa-Ichiki model $\frac{(1-\phi)^3}{1+2\phi+1.429\phi (1-\phi)^3}$ \citep{hayakawa1995statistical}. The values chosen for the exponent $n$ and the coefficient $S$ here are typically for non-Brownian particles interacting only hydrodynamically \citep{padding2004hydrodynamic,moncho2010effects}.

Our simulation results agree with Batchelor’s model for $\phi$ approximately smaller than 0.03. For larger volume fractions, the simulation gives larger values than Batchelor’s model. A similar range of validity for Batchelor’s model was also found by \citet{abbas2006dynamics} using a force-coupling method. Our results also agree well with the Hayakawa-Ichiki model for $\phi \leq 0.05$ and they lie between the predictions of Richardson-Zaki's correlation and Hayakawa-Ichiki's model for $\phi \geq 0.06$. The simulation data for $\phi=0.01$ is smaller than the values predicted by the three models. This is expected because of the use of triply-periodic boundary conditions in a domain of finite size \citep{phillips1988hydrodynamic}.

\begin{figure}
  \centerline{\includegraphics[width=1.0\textwidth]{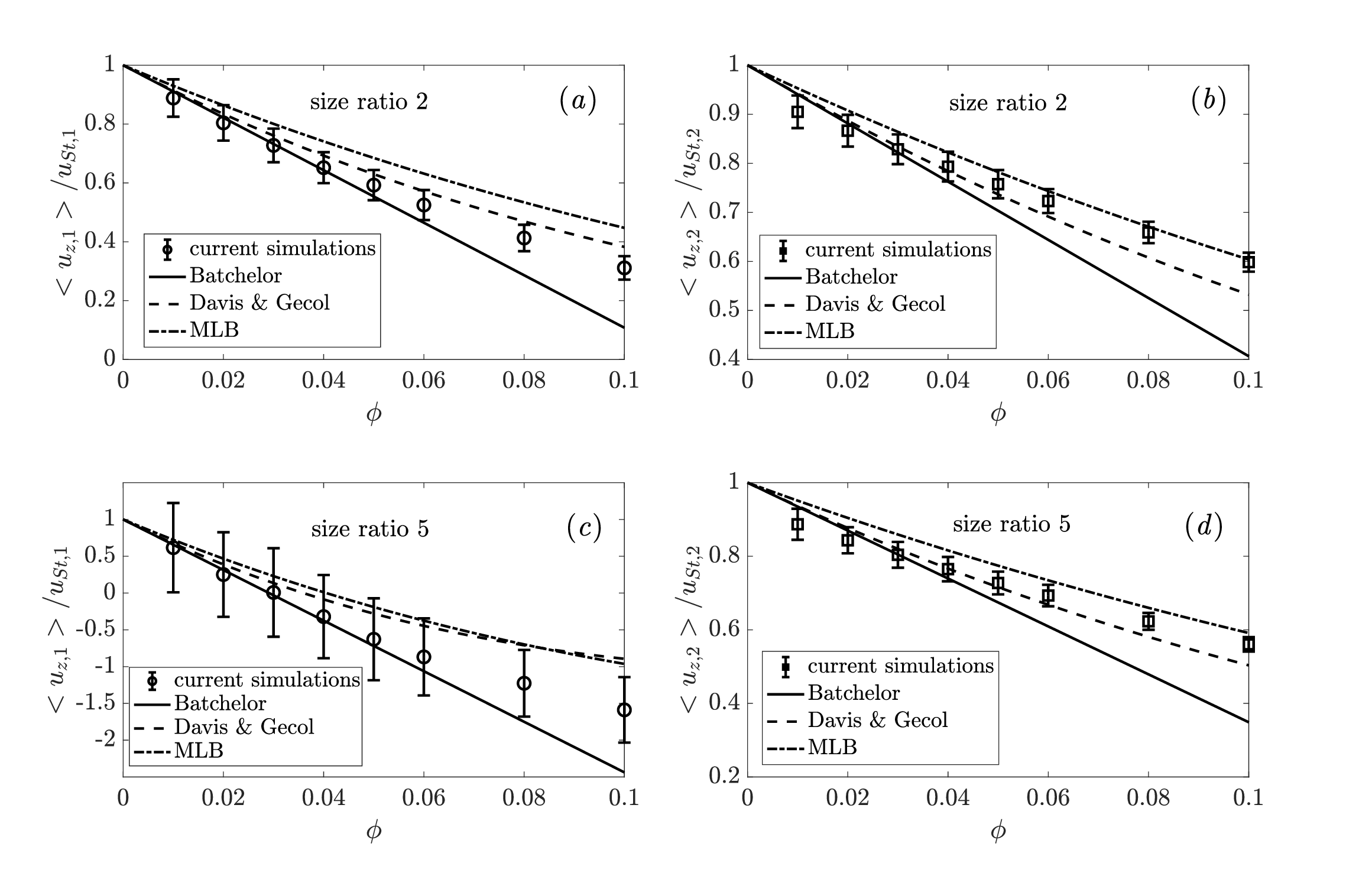}}% Images in 100% size
  \caption{Bidisperse case: average settling velocity normalized by the single-particle Stokes velocities for the small (left panels) and large (right panels) particles. Panels (\textit{a}) and (\textit{b}) are for size ratio 2. Panels (\textit{c}) and (\textit{d}) are for size ratio 5.}
\label{fig:bi_mean}
\end{figure}

Normalised average settling velocities for the small and the large particles in the bidisperse case are plotted as symbols in figure \ref{fig:bi_mean} for two size ratios. The predictions of Batchelor's model (equation (\ref{batchelor})), Davis \& Gecol's model (equation (\ref{dg})) and the MLB model (equation (\ref{mlb})) are indicated by lines. It is seen from figure \ref{fig:bi_mean} (\textit{a}) and (\textit{c}) that our results for the small particles agree with the predictions of Batchelor's model for $\phi \leq 0.05$, and lie between the predictions from Batchelor's model and Davis \& Gecol's models for $\phi \geq 0.06$. For the large particles, our results agree with predictions from Batchelor's model for $\phi \leq 0.03$ and lie between the predictions from Davis \& Gecol's and MLB models for $\phi \geq 0.04$.  Stokesian dynamics calculations by \citet{wang2015short} that include stresslet and lubrication contributions also predicted for $\phi$ larger than around 0.05 hindered settling velocities smaller and larger than those of Davis \& Gecol's for the small and the large particles, respectively.  

In summary our simulation results for the mono- and bi-disperse cases are comparable to those in the literature. 

\section{Polydisperse suspensions}
\label{sec:poly}

\begin{figure}
  \centerline{\includegraphics[width=1.0\textwidth]{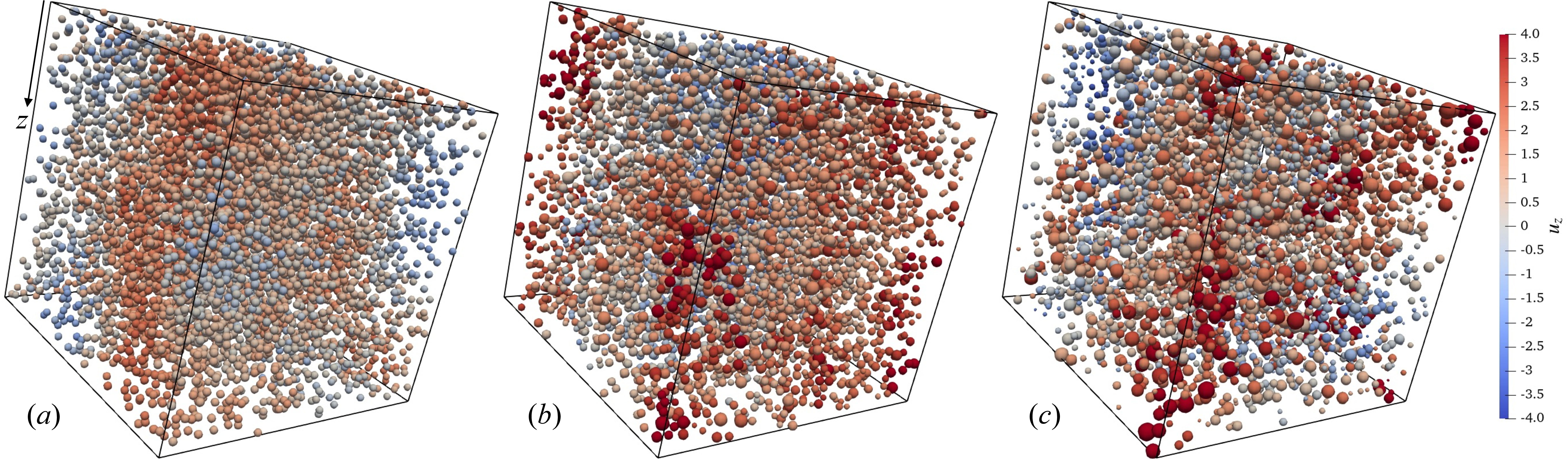}}% Images in 100% size
  \caption{Configuration for different polydispersity parameters and $\phi$=0.05, with the spheres coloured according to their settling velocity; (\textit{a}) $\alpha$=0, (\textit{b}) $\alpha$=0.2, and (\textit{c}) $\alpha$=0.4.}
\label{fig:particle_velocity_snapshot}
\end{figure}
To illustrate the spatial distribution of particle velocities in the polydisperse particle simulations, in figure \ref{fig:particle_velocity_snapshot} we show snapshots of the simulations with each sphere coloured according to its settling velocity. Spheres coloured in red have settling velocities in the direction of gravity whereas spheres coloured in blue have settling velocities opposite to gravity. Figure \ref{fig:particle_velocity_snapshot} (\textit{c}) shows that the smaller particles can move against gravity, and have negative velocities comparable in magnitude to the positive settling velocity of the largest particles.

\begin{figure}
  \centerline{\includegraphics[width=1.0\textwidth]{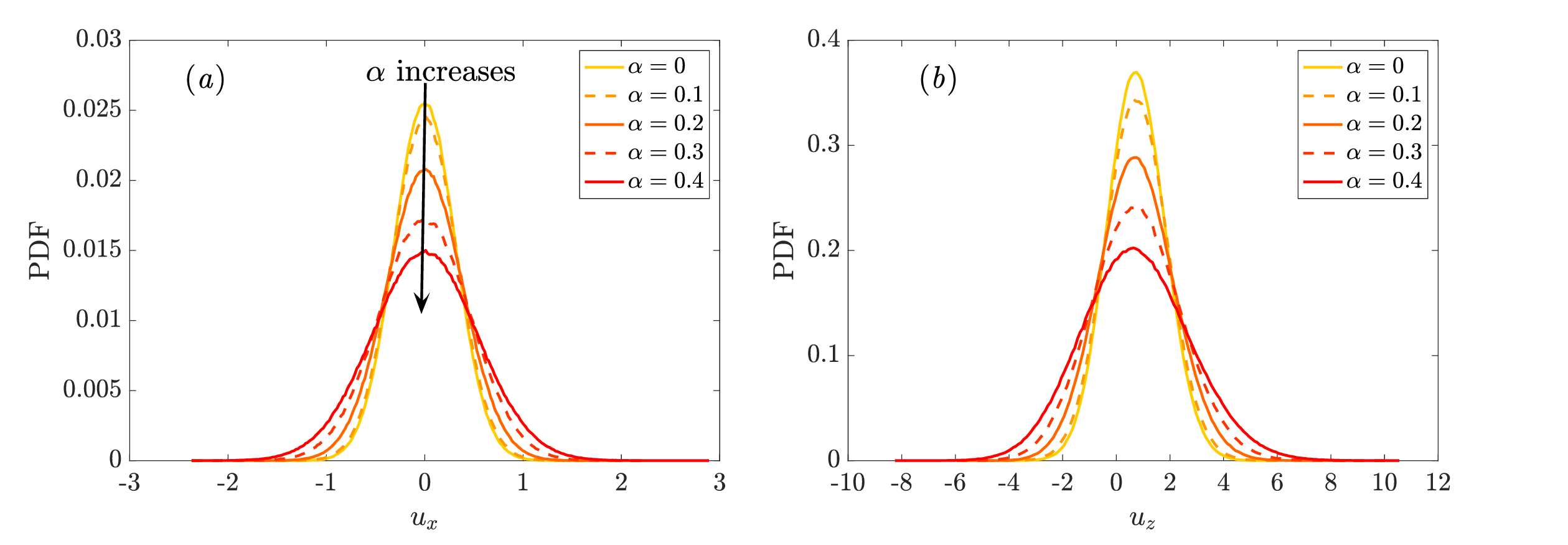}}% Images in 100% size
  \caption{Probability distribution functions of (\textit{a})  horizontal\protect\\ and (\textit{b}) vertical velocities for different polydispersity parameters and $\phi$=0.05.}
\label{fig:pdf_all}
\end{figure}

\begin{figure}
  \centerline{\includegraphics[width=1.0\textwidth]{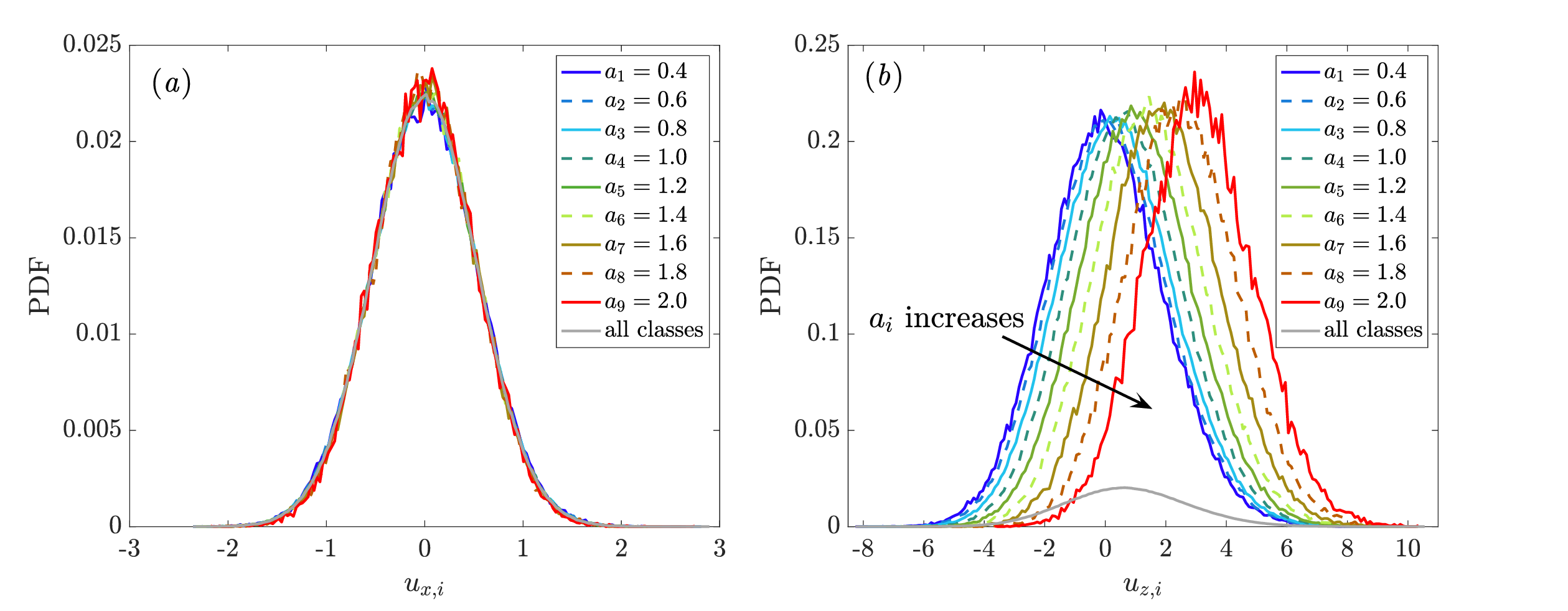}}% Images in 100% size
  \caption{Probability distribution functions of (\textit{a})  horizontal\protect\\ and (\textit{b}) for $\alpha$=0.4 at $\phi$=0.05. In contrast to Fig. \ref{fig:pdf_all}, here the PDFs are calculated based on the distribution of velocities within each size class.}
\label{fig:pdf_each}
\end{figure}

Probability distribution functions (PDFs) of horizontal and vertical velocities,  shown in figure \ref{fig:pdf_all} for different values of $\alpha$, are approximately Gaussian, with a variance that increases as $\alpha$ increases. These PDFs are constructed by considering all the particles in the simulation domain. However, spheres belonging to the same size class also have a distribution of settling velocities. Therefore, in figure \ref{fig:pdf_each}, we show the PDFs of the horizontal and the vertical velocities of spheres in each size class for $\alpha$=0.4. For comparison, the PDFs of the velocities of all the spheres are included in this plot as grey lines. Again, the PDFs are approximately Gaussian (simulations by \citet{cheng2023hydrodynamic} of uniform flow past fixed polydisperse random arrays indicate also a Gaussian distribution for the hydrodynamic forces of a given size class). Surprisingly, the PDFs of the horizontal velocities for different size classes collapse onto a single curve (figure \ref{fig:pdf_each} (\textit{a})). From the PDFs of the vertical velocities in \ref{fig:pdf_each} (\textit{b}), it is seen that the mean velocity increases as the size of the particle increases. And, different size classes have comparable variances. 

\begin{figure}
  \centerline{\includegraphics[width=0.6\textwidth]{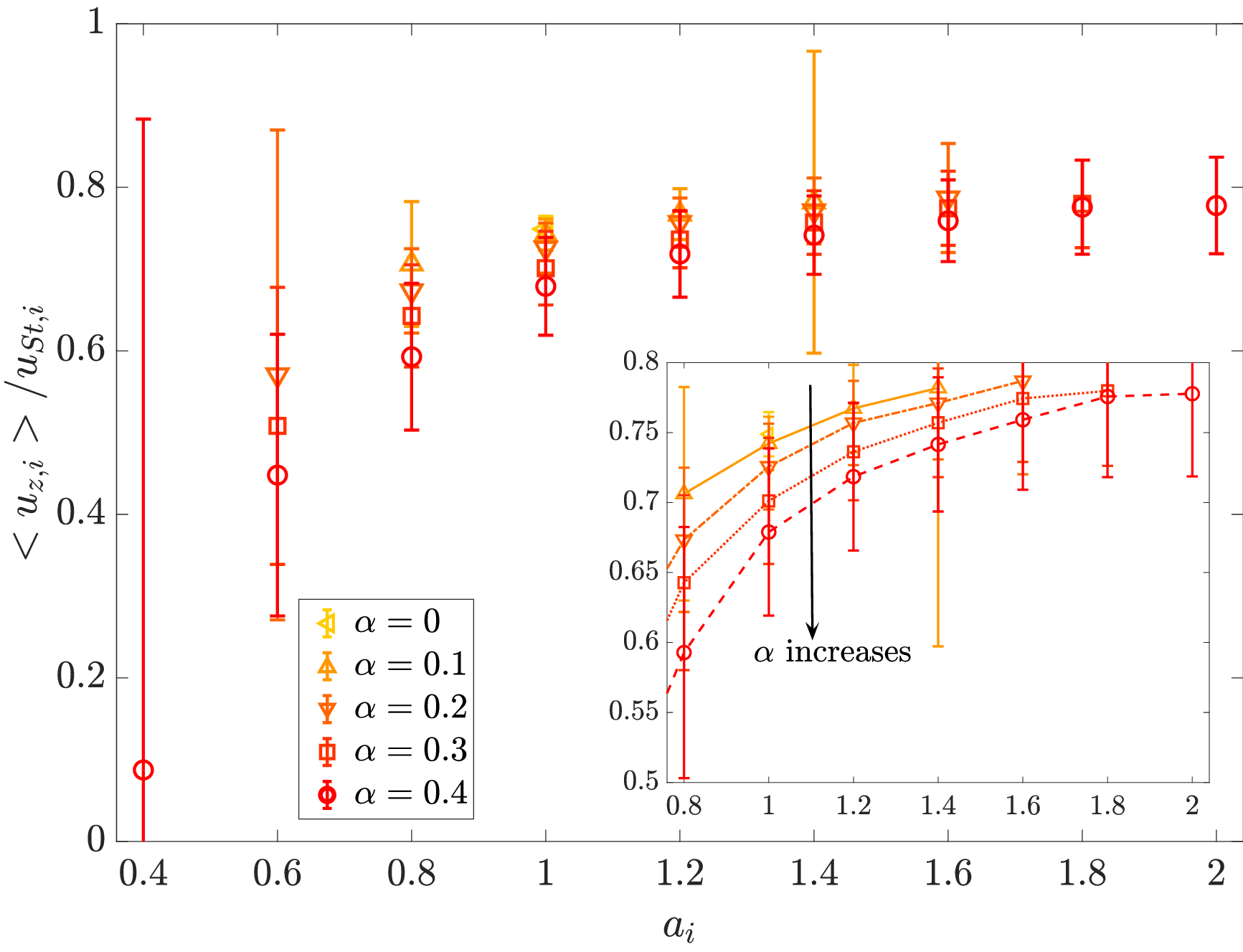}}% Images in 100% size
  \caption{Normalized average settling velocity of each size class for different polydispersity parameters and $\phi$=0.05. The inset is a zoom in the range $0.8 \leq a_i \leq 2$. The lines are guides for the eyes.}
\label{fig:average_polydispersity}
\end{figure}

The average vertical settling velocity of each size class normalized by the corresponding single-particle Stokes velocity is shown in figure \ref{fig:average_polydispersity} for different values of $\alpha$. The inset  shows a zoom in the range $0.8 \leq a_i \leq 2$. Because now the settling velocity is normalised by the single-particle settling velocity, the information in this plot complements the data of figure \ref{fig:pdf_each} (\textit{b}). We see that for fixed $\alpha$ the normalized average settling velocity increases as the particle size increases. This means that the velocities of small particles are more hindered than the velocities of large particles. For a given size class, the normalized average settling velocity decreases as $\alpha$ increases, and decreases faster for small particles than for large particles.

\begin{figure}
  \centerline{\includegraphics[width=0.6\textwidth]{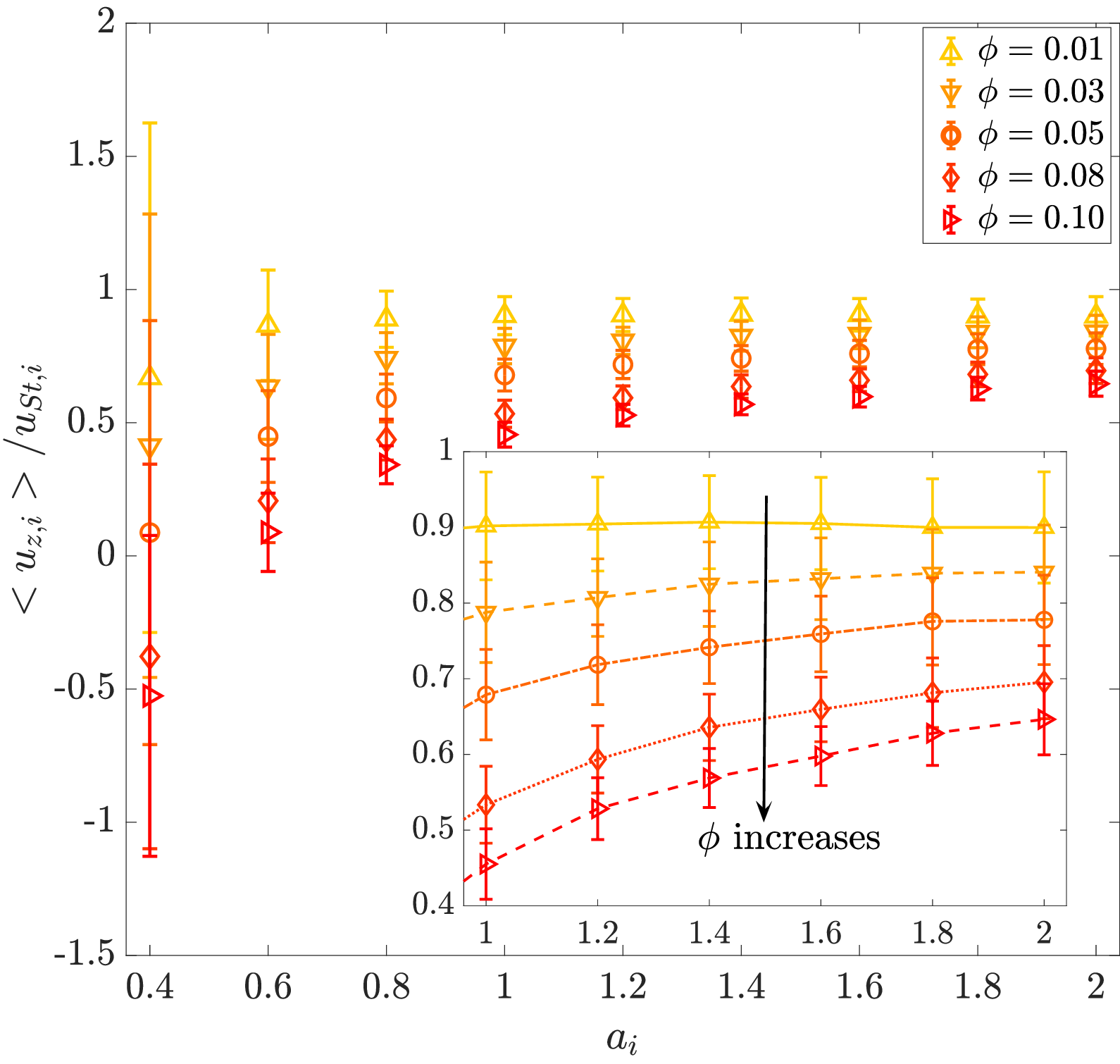}}% Images in 100% size
  \caption{Normalized average settling velocities of each size class, for $\alpha$=0.4 and different volume fractions. The inset shows a zoom in the range $1 \leq a_i \leq 2$.}
\label{fig:average_vf}
\end{figure}

In the previous figures, the total volume fraction was fixed, and $\alpha$ was changed. In figure \ref{fig:average_vf}, we instead change $\phi$ for fixed $\alpha$=0.4. This plot confirms the trend seen in figure \ref{fig:average_polydispersity}: for a given volume fraction, the normalized average settling velocity decreases as the particle size decreases. The normalized average settling velocity decreases faster with increasing $\phi$ for small size particles. 

To summarise, the smaller particles are more hindered and more affected by polydispersity than the large ones. 

\subsection{Comparison with hindered settling models}
\label{sec:comparison}

In this subsection, current simulations are compared with predictions from Batchelor's (see equation (\ref{batchelor})), Davis \& Gecol's (see equation (\ref{dg})) and MLB (see equation (\ref{mlb})) models. The accuracy of Richardson-Zaki correlation (see equation (\ref{r-z})) for polydisperse suspensions is also checked. The values of the coefficients $S_{ij}$ in Batchelor's and Davis \& Gecol's models are calculated from $S_{ij}=-3.50-1.10\lambda-1.02\lambda^2-0.002\lambda^3$ where $\lambda=a_j/a_i$ \citep{davis1994hindered}. The value of the exponent $n$ in the MLB model and the Richardson-Zaki correlation is 5.

\begin{figure}
  \centerline{\includegraphics[width=1.0\textwidth]{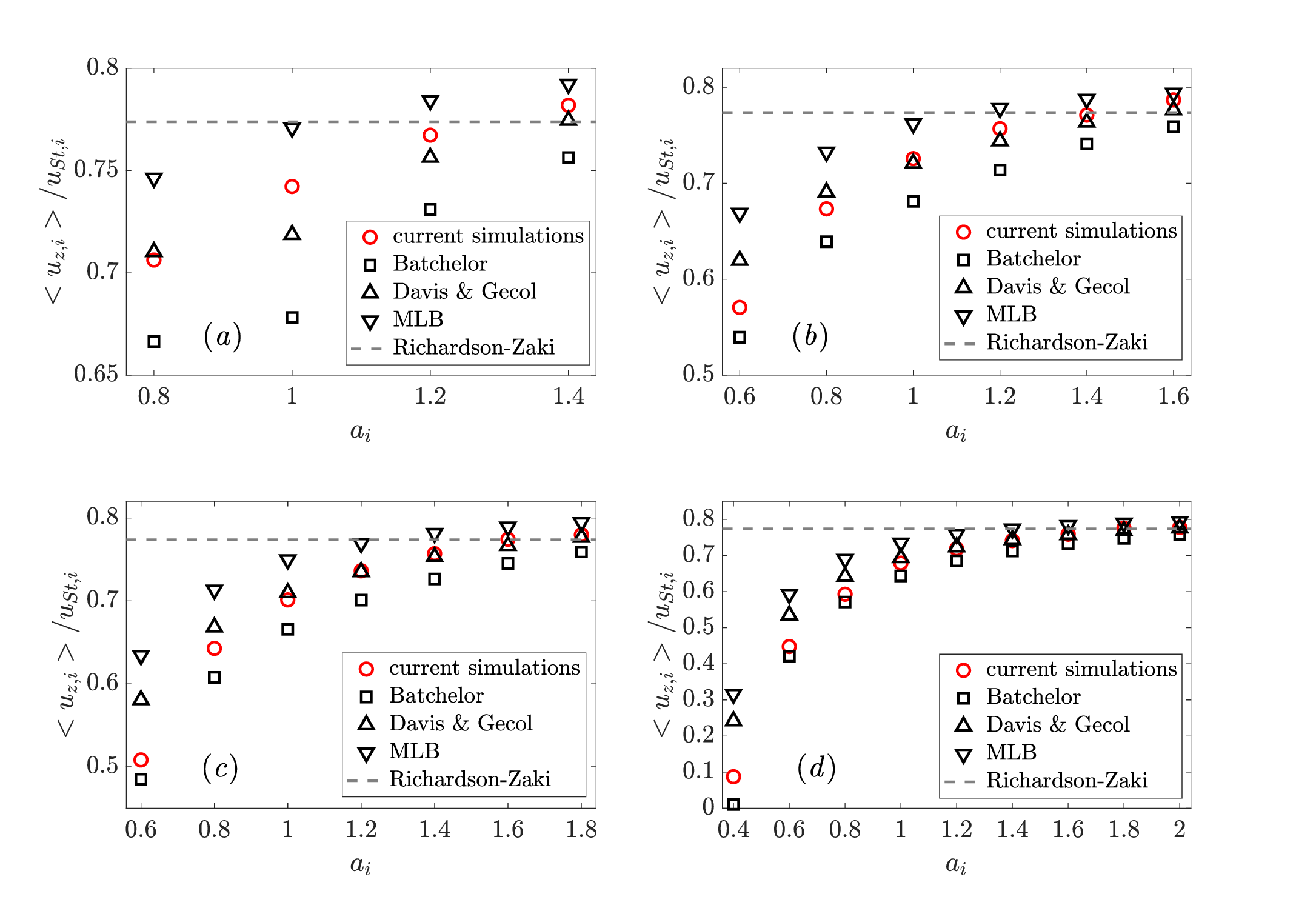}}% Images in 100% size
  \caption{Comparison between the current simulation results  and the predictions of hindered settling function models for the average velocity of different size classes, for $\phi$=0.05 and different polydispersity parameters: (\textit{a}) $\alpha$=0.1, (\textit{b}) $\alpha$=0.2, (\textit{c}) $\alpha$=0.3 and (\textit{d}) $\alpha$=0.4.}
\label{fig:compare_poly}
\end{figure}

Hindered settling functions corresponding to different size classes for fixed $\phi$=0.05 and different $\alpha$ are compared against different theoretical models in figure \ref{fig:compare_poly}. The Richardson-Zaki correlation largely overestimates the hindered settling functions of smaller particles, whereas it gives reasonable values for larger particles. The discrepancy between the Richardson-Zaki correlation and the computed hindered settling functions of smaller particles increases as $\alpha$ increases. For each $\alpha$, the predictions from the other three models show a consistent trend for each size class. The MLB model gives the largest settling velocities, Batchelor's model gives the smallest settling velocities, and Davis \& Gecol's model gives intermediate values. The differences between the predictions from these three models get smaller as the particle size increases. 

\begin{figure}
  \centerline{\includegraphics[width=0.6\textwidth]{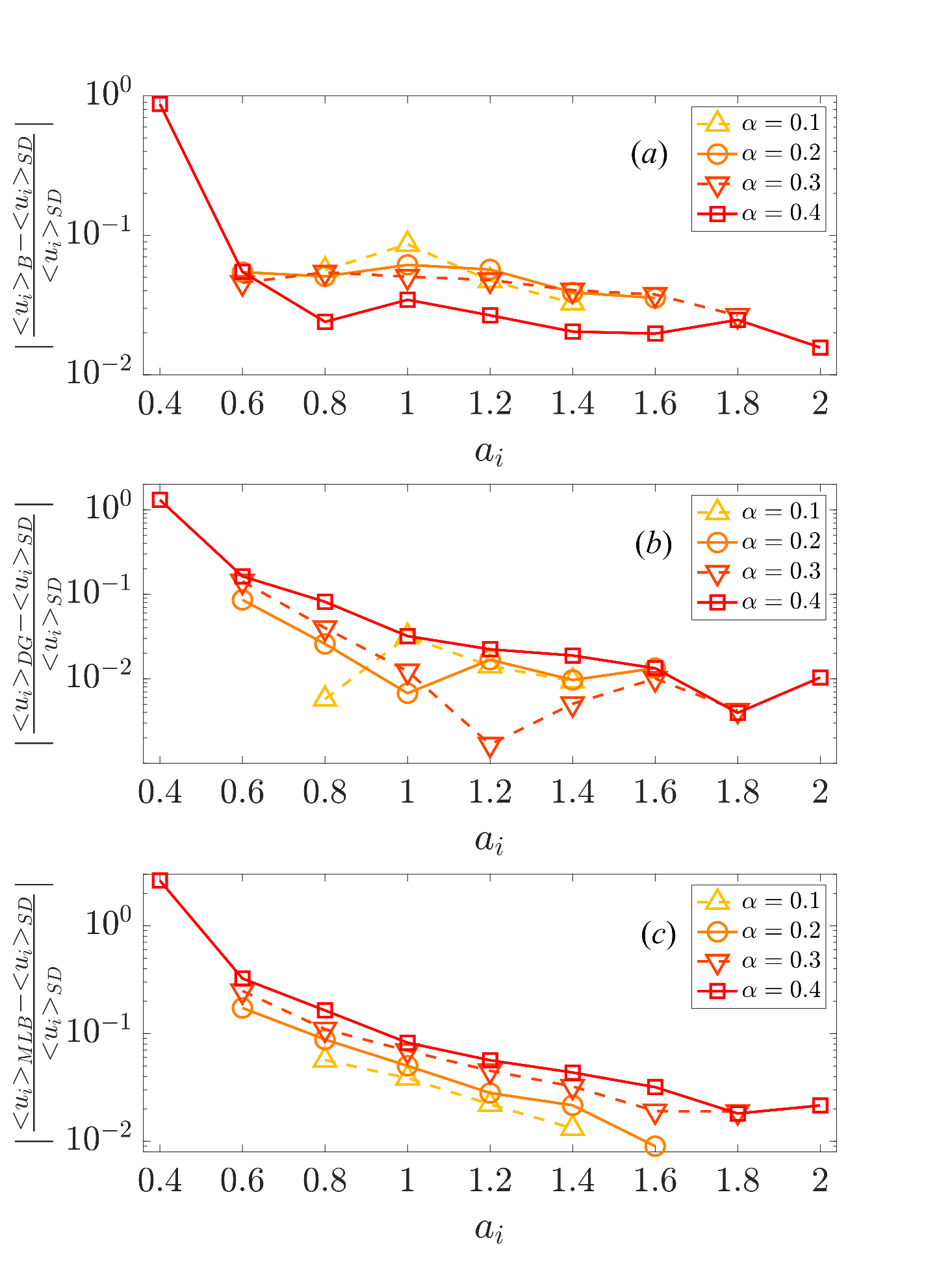}}% Images in 100% size
  \caption{Relative differences between the average settling velocities from different models and from the current simulations for fixed $\phi = 0.05$: (\textit{a}) Batchelor's model; (\textit{b}) Davis \& Gecol model;  (\textit{c}) MLB model.}
\label{fig:rela_poly}
\end{figure}

Figure \ref{fig:rela_poly} shows the normalized relative differences between the computed and predicted average settling velocities. The Batchelor model and the Davis \& Gecol model predict the average settling velocity of each size class quite well for all $\alpha$ considered here, with relative errors smaller than 10\%, except for the smallest size class $a_i$=0.4 for $\alpha$=0.4 for which the simulation gives a very small settling velocity. From figure \ref{fig:rela_poly} (\textit{c}), it is seen that the relative difference between the predictions from the MLB model and the current simulations decreases as $a_i$ increases, or $\alpha$ decreases. For $a_i \geq 1$, the MLB model predicts the average settling velocities quite well, with the relative difference within 10\%. For $a_i \leq 0.8$, the MLB model starts failing.

\begin{figure}
  \centerline{\includegraphics[width=1.0\textwidth]{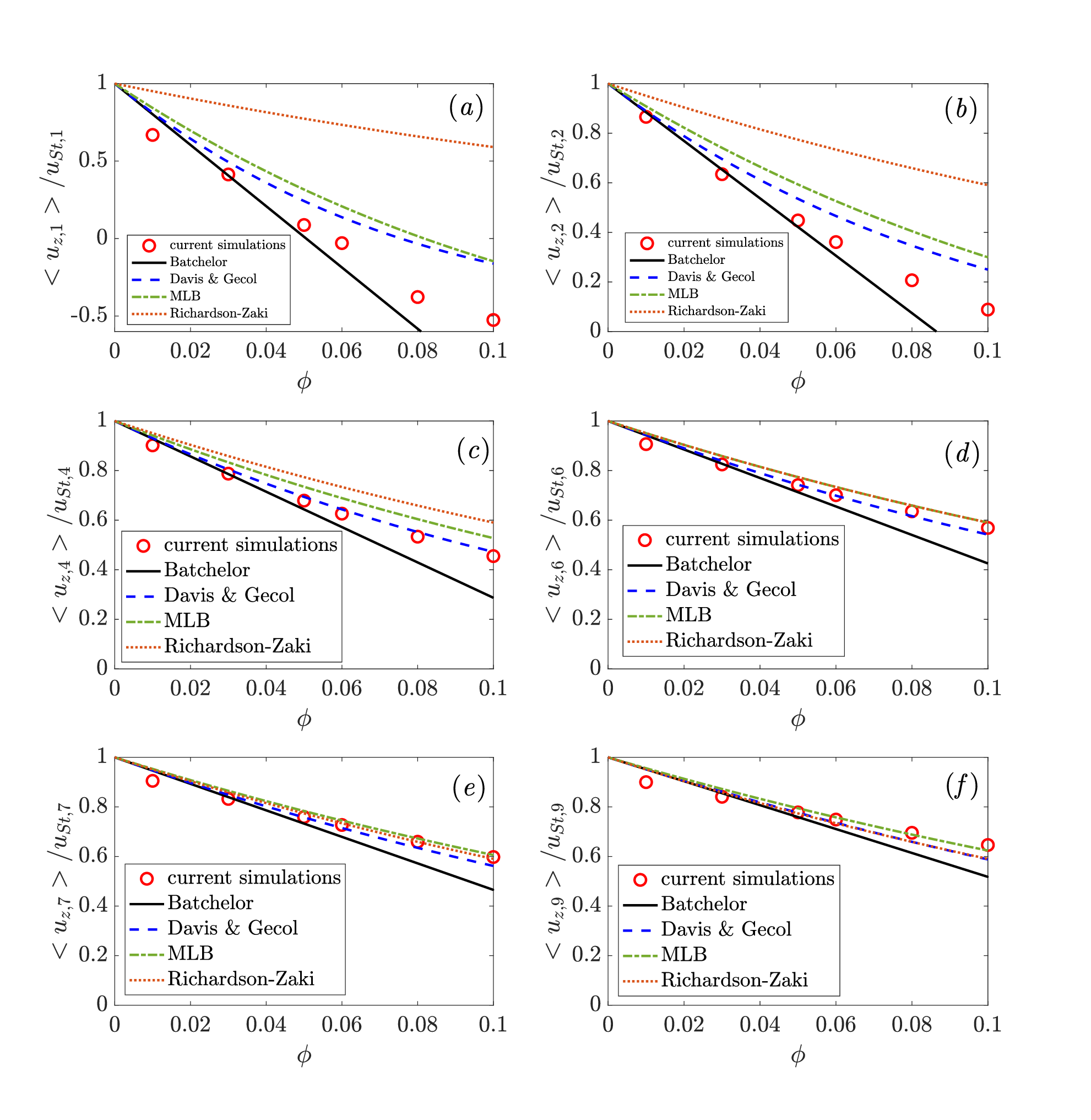}}% Images in 100% size
  \caption{Comparison between the current numerical results and the predictions of different models. The comparison is here evaluated as a function of $\phi$  for fixed $\alpha$=0.4 . (\textit{a}) to (\textit{f}) correspond to size classes $a_1$,$a_2$,$a_4$,$a_6$,$a_7$ and $a_9$, respectively.}
\label{fig:compare_vf}
\end{figure}

Hindered settling function data for fixed $\alpha$=0.4 and different $\phi$ are compared with the predictions from different models in figure \ref{fig:compare_vf}. Also, Richardson-Zaki's correlation predicts the hindered settling functions of smaller particles poorly. For the other three models, a similar trend in the predicted values is observed as the one in the case of varying $\alpha$. The predictions from the MLB model and the Davis \& Gecol model are quite close to each other for all size classes at each volume fraction, and they are also quite close to the values from current simulations for larger particles. However, the Davis \& Gecol model slightly underestimates the hindered settling functions of the larger particles when $\phi > 0.06$. For smaller particles, the predictions from the MLB model and the Davis \& Gecol model are larger than the values from the simulations, and the discrepancies between the predictions from these two models and the values from current simulations get larger as $\phi$ increases. The predictions of Batchelor's model are close to the simulated values for $\phi$ approximately less than 0.05. As $\phi$ increases, Batchelor's model underestimates the hindered settling functions of all size classes systematically compared to the results of current simulations.

\begin{figure}
  \centerline{\includegraphics[width=0.6\textwidth]{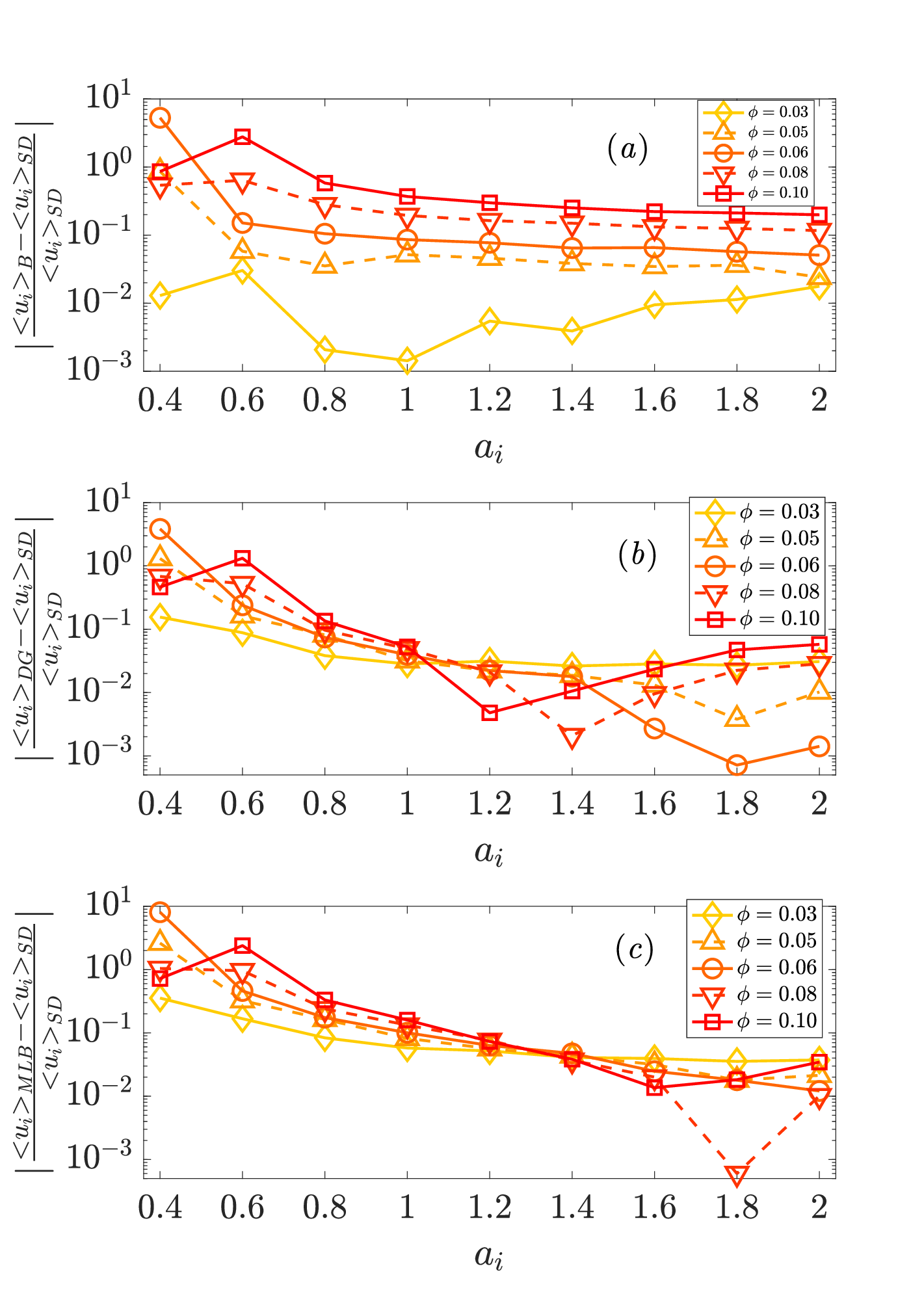}}% Images in 100% size
  \caption{Same as in Fig. \ref{fig:rela_poly}, but for different volume fractions and fixed $\alpha = 0.4.$}
\label{fig:rela_vf}
\end{figure}

For fixed $\alpha$=0.4, the relative differences between the average settling velocities predicted by different models and calculated by current simulations of each size class for different volume fractions are shown in figure \ref{fig:rela_vf}. For each size class, the relative difference between the prediction from the Batchelor model and the current simulations increases as the volume fraction increases, and it is within 10\% when $\phi \leq 0.05$, except for the smallest size class $a_i=0.4$. From figure \ref{fig:rela_vf} (\textit{b}) and (\textit{c}), it is seen that the relative differences are quite close for the Davis \& Gecol and the MLB models, with those of the MLB model slightly larger. For larger size classes ($a_i\geq 1$), both these two models give quite accurate predictions for all volume fractions considered, with the relative differences within 10\% compared to the results of current simulations. For smaller size classes ($a_i \leq 0.8$), both these two models give predictions with large relative differences compared to the results of current simulations, and in general the relative difference gets larger as volume fraction increases or as size of the class decreases.

We saw that the MLB model, despite its simplicity, gives relatively good agreement for the large particles. However, it fails for the small particles. The MLB model is based on a closure relation for the particle-fluid velocity difference (see Appendix \ref{appA}). Therefore, to understand the limits of validity of the model, we compute the slip velocity from the simulation data and compare against the MLB model prediction. 

\begin{figure}
  \centerline{\includegraphics[width=0.6\textwidth]{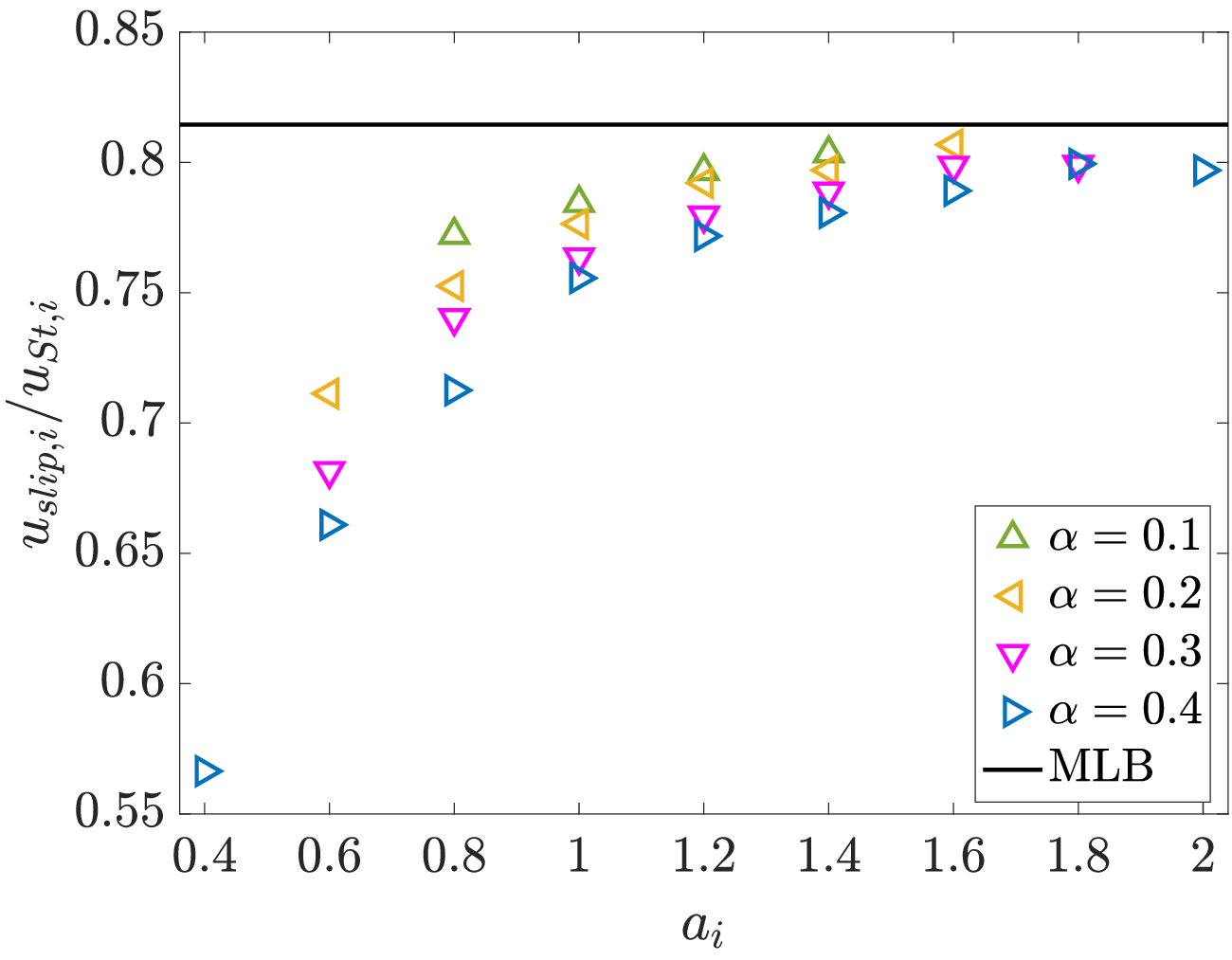}}% Images in 100% size
  \caption{Normalized slip velocity for each size class for $\phi$=0.05 and different $\alpha$. The line shows the prediction of the MLB model.}
\label{fig:slip_poly}
\end{figure}

Slip velocities for each size class normalized by the corresponding Stokes velocities are plotted in figure \ref{fig:slip_poly} for $\phi$=0.05 and different $\alpha$. The slip velocity for size class $i$ is defined as the difference between the average settling velocity of class $i$ and the average velocity of the fluid phase, 
\begin{equation}
u_{slip,i}=\langle u_{z,i} \rangle - \langle u_f \rangle.
\label{slipdefinition}
\end{equation}
The value of $\langle u_f \rangle$ is obtained from the the zero volume-flux condition $\sum\phi_{j} \langle u_{z,j} \rangle + (1-\phi) \langle u_f \rangle = 0$. The slip velocity predicted by the MLB model is calculated from (see Appendix \ref{appA}) 
\begin{equation}
u_{slip,i}=u_{St,i}(1-\phi)^{n-1}. 
\end{equation}

It is seen that the MLB model does not predict accurately the slip velocities of the smaller particles. The discrepancy between the MLB model prediction and the simulation data increases as $\alpha$ increases. The slip velocities of relatively large particles are reasonably well captured. As the particle size increases, the simulation data tends to converge to the MLB model prediction. 

\begin{figure}
  \centerline{\includegraphics[width=0.6\textwidth]{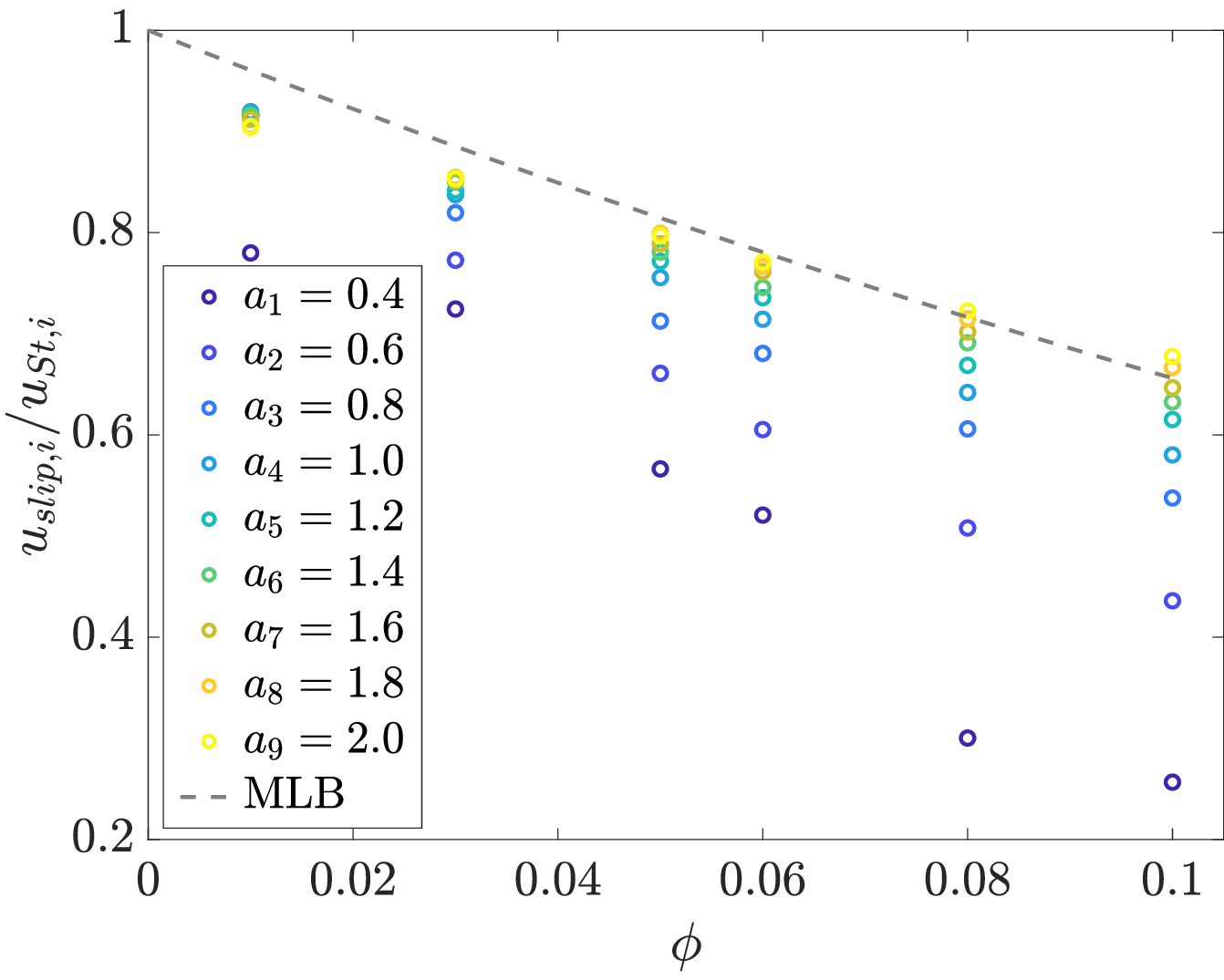}}% Images in 100% size
  \caption{Normalized slip velocity for each size class for $\alpha$=0.4 and different values of $\phi$. The dashed line is the prediction of the MLB model.}
\label{fig:slip_vf}
\end{figure}

The normalized slip velocities for each size class for fixed $\alpha$=0.4 and varying $\phi$ are plotted in figure \ref{fig:slip_vf}. It is seen that the prediction of the MLB model gets increasingly worse as the volume fraction increases for the small size classes. Predictions for the largest particles are instead acceptable regardless of the volume fraction.

\section{Particle velocity fluctuations}
\label{sec:fluctuations}

\begin{figure}
  \centerline{\includegraphics[width=1.0\textwidth]{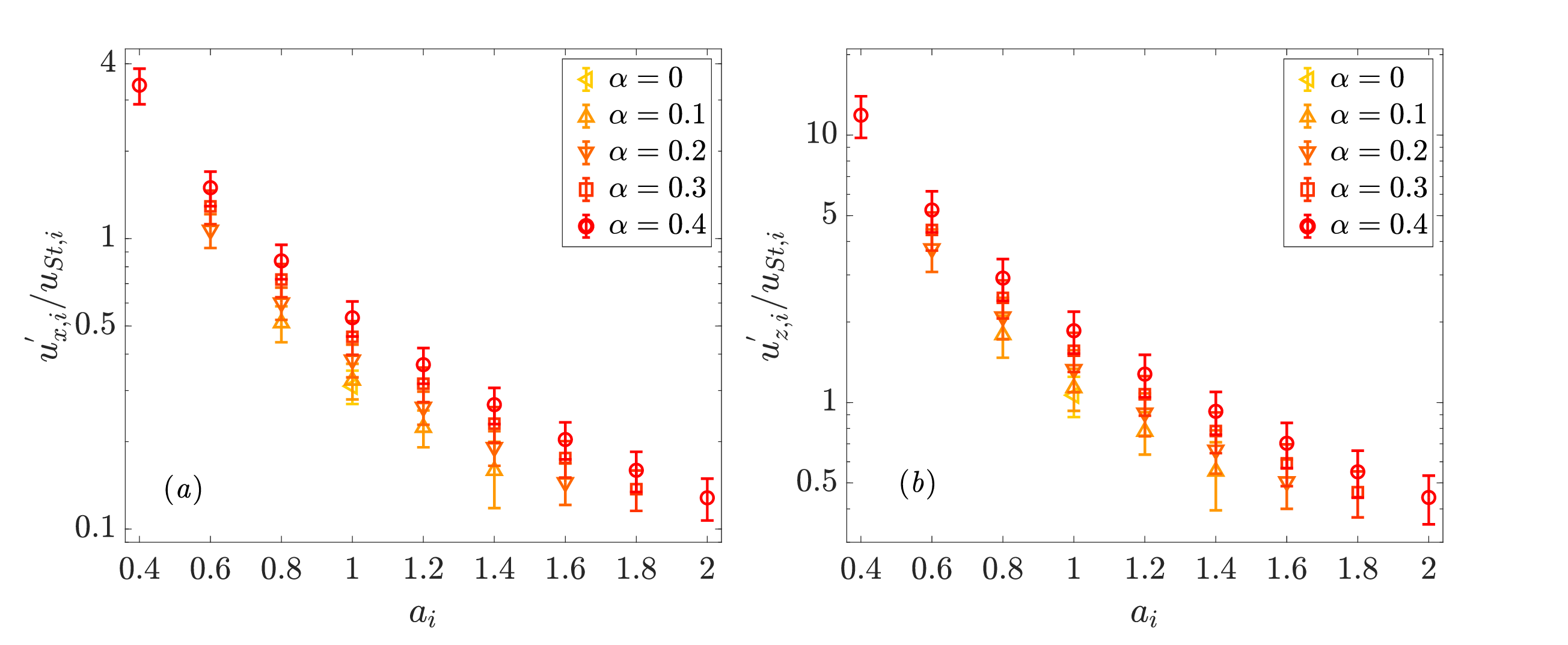}}% Images in 100% size
  \caption{Normalized  (\textit{a}) horizontal\protect\\ and (\textit{b})  vertical velocity fluctuations for different $\alpha$ and $\phi$=0.05.}
\label{fig:fluc_poly_poly}
\end{figure}

\begin{figure}
  \centerline{\includegraphics[width=1.0\textwidth]{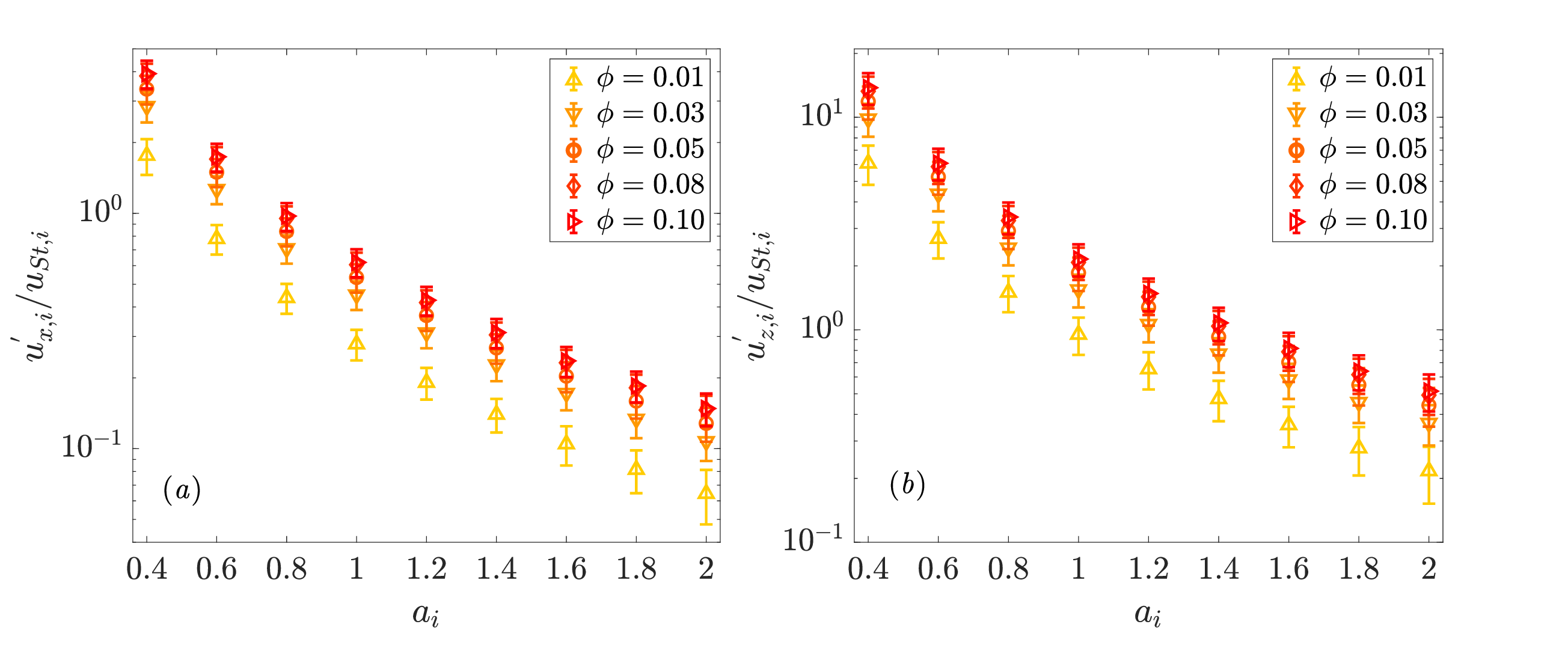}}% Images in 100% size
  \caption{Normalized (\textit{a})  horizontal\protect\\ and (\textit{b})  vertical velocity fluctuations for different $\phi$ and $\alpha$=0.4.}
\label{fig:fluc_poly_vf}
\end{figure}

\begin{figure}
  \centerline{\includegraphics[width=1.0\textwidth]{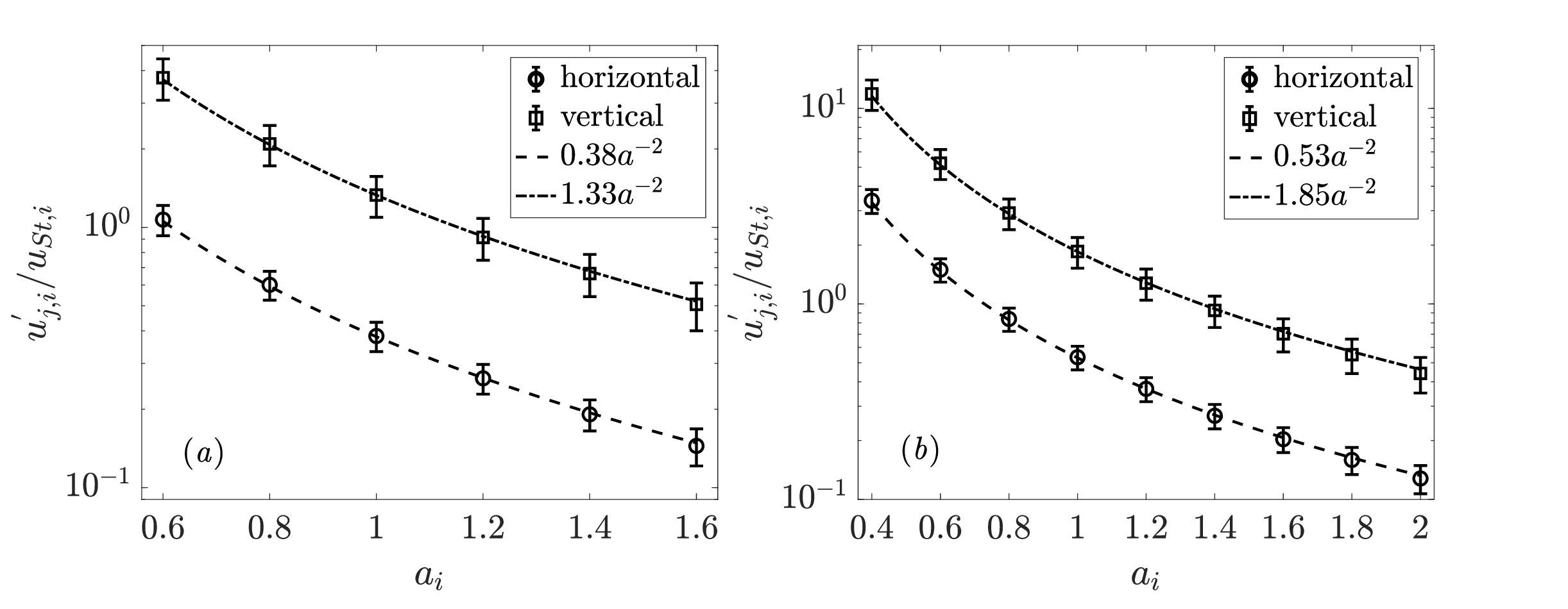}}% Images in 100% size
  \caption{Normalized velocity fluctuations of each size class at $\phi=0.05$ for (\textit{a}) $\alpha=0.2$\protect\\ and (\textit{b}) $\alpha=0.4$. }
\label{fig:fluc_poly}
\end{figure}

Statistical deviations with respect to the mean particle velocity, as measured by the root-mean square of the velocity fluctuation,  increase as $\alpha$ or $\phi$ increase (see figures \ref{fig:fluc_poly_poly} and \ref{fig:fluc_poly_vf}). The normalized horizontal and vertical velocity fluctuations of each size class are shown for $\alpha = 0.2$ and 0.4 with fixed $\phi=0.05$ in figure \ref{fig:fluc_poly}. For a fixed $\alpha$, the normalized velocity fluctuations decrease as the particle size increases. Figure \ref{fig:pdf_each} seems to suggest that the velocity fluctuations are approximately independent of the particle radius $a_i$. Because the Stokes velocity scales as $a_i^2$, it is expected that the velocity fluctuations normalized by the Stokes velocity scale as $\sim a_i^{-2}$. Our data confirm this scaling  (see lines in figure \ref{fig:fluc_poly}): $u_i^{\prime}/u_{St,i}=ca_i^{-2}$ fits the data for all the values of $\alpha$ and $\phi$ simulated, as shown in tables \ref{tab:poly} and \ref{tab:vf} (this scaling is also observed in our simulations of bidisperse suspensions, as velocity fluctuations for the two classes are similar in magnitude).

\begin{table}
  \begin{center}
\def~{\hphantom{0}}
  \begin{tabular}{lcc}
      $\alpha$ & horizontal direction  & vertical direction  \\[3pt]
       0.1   & 0.32 & 1.13 \\
       0.2   & 0.38 & 1.33 \\
       0.3   & 0.46 & 1.56 \\
       0.4   & 0.53 & 1.85 \\
  \end{tabular}
  \caption{Approximate values of the prefactor $c$ in the scalings of the horizontal and the vertical normalized velocity fluctuations for each $\alpha$ at $\phi=0.05$.}
  \label{tab:poly}
  \end{center}
\end{table}

\begin{table}
  \begin{center}
\def~{\hphantom{0}}
  \begin{tabular}{lcc}
      $\phi$ & horizontal direction  & vertical direction  \\[3pt]
      0.01   & 0.28 & 0.95 \\
       0.03   & 0.45 & 1.54 \\
       0.05   & 0.53 & 1.85 \\
       0.08   & 0.61 & 2.08 \\
       0.10   & 0.62 & 2.15 \\
  \end{tabular}
  \caption{Same as table \ref{tab:poly} but for different $\phi$ at $\alpha=0.4$.}
  \label{tab:vf}
  \end{center}
\end{table}

\begin{figure}
  \centerline{\includegraphics[width=1.0\textwidth]{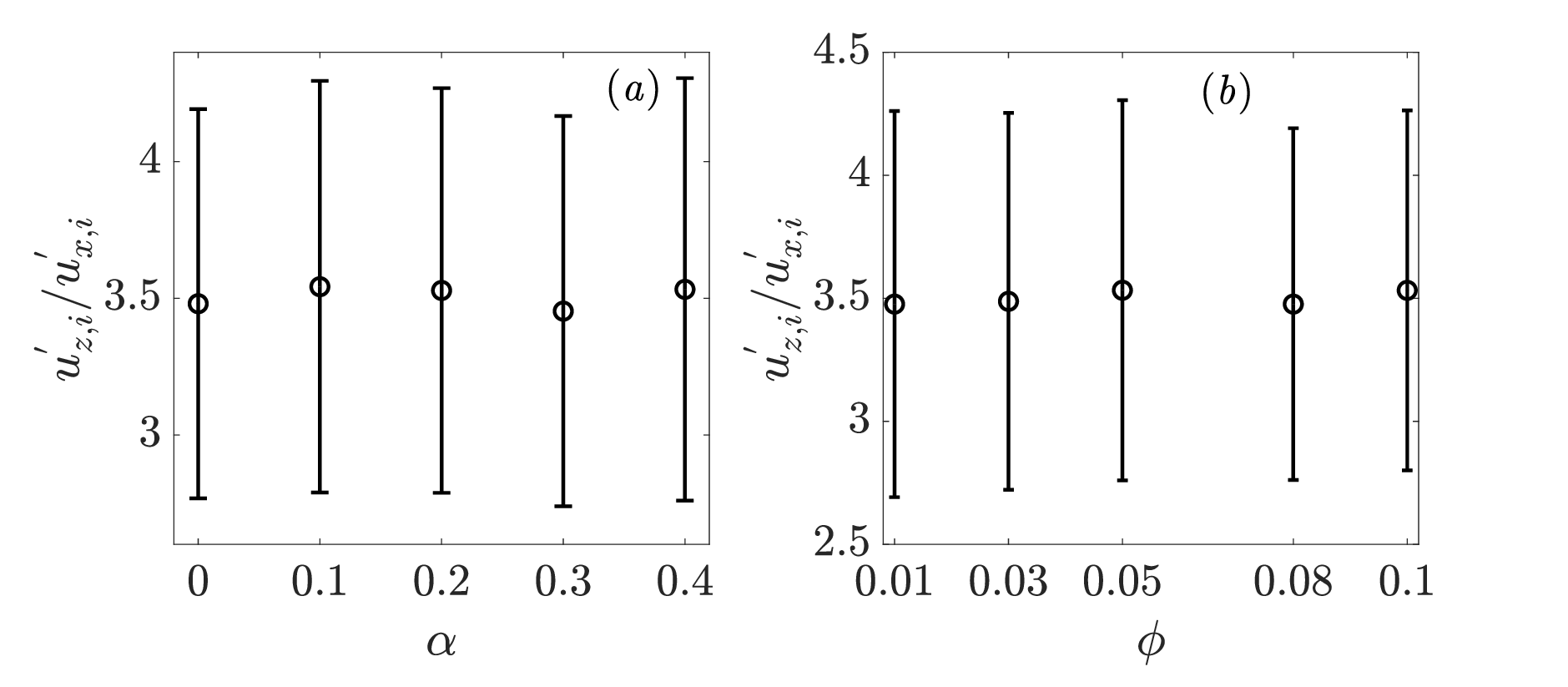}}% Images in 100% size
  \caption{Ratio between  vertical and  horizontal velocity fluctuation magnitudes for $a_i=1$ and (\textit{a}) different $\alpha$ at $\phi=0.05$,\protect\\ or (\textit{b}) different $\phi$ at $\alpha=0.4$.}
\label{fig:fluc_poly_anis}
\end{figure}

 The anisotropy ratio between the vertical and the horizontal velocity fluctuations,  plotted in figure \ref{fig:fluc_poly_anis},  is around 3.5 regardless of the values of $\alpha$ or $\phi$. This value was also observed in the monodisperse and bidisperse simulations. 
 
% For settling polydisperse suspensions, very few works discussed about velocity fluctuations, and all of them were for bidisperse systems \citep{peysson1999velocity,zaidi2020settling}. 

\citet{peysson1999velocity} measured experimentally the velocity fluctuations of  small and  large particles in a dilute bidisperse suspensions with size ratio 2. They found that the ratio of velocity fluctuations between the small and the large size classes were around 0.85 and 0.75 in the vertical and horizontal directions, respectively, close to $(a_1/a_2)^{1/2}\approx 0.71$. To explain their findings, they extended Hinch’s scaling for the velocity fluctuations in monodisperse suspensions \citep{hinch1988sedimentation}, using two different correlation lengths for the two size classes. In our simulations, the correlation length should be the size of the computational box, as found in other numerical simulations using periodic boxes \citep{ladd1996hydrodynamic,nguyen2005sedimentation} and in other experiments during the initial settling stage of well-mixed suspensions \citep{segre1997long,tee2002nonuniversal}. Using  the same correlation length for the two size classes would predict comparable velocity fluctuation magnitudes.  
%%%%%%%%%%%%%%%%%%%%%%%%%%%%%%%%%%%%%%%%%%%%%%%%%%%%%%%%
\section{Discussion}
\label{sec:discussion}

% \subsection{Batchelor's model and its extension}
\label{sub:batchelor}
\begin{figure}
    \centering
    \includegraphics[width=0.6\textwidth]{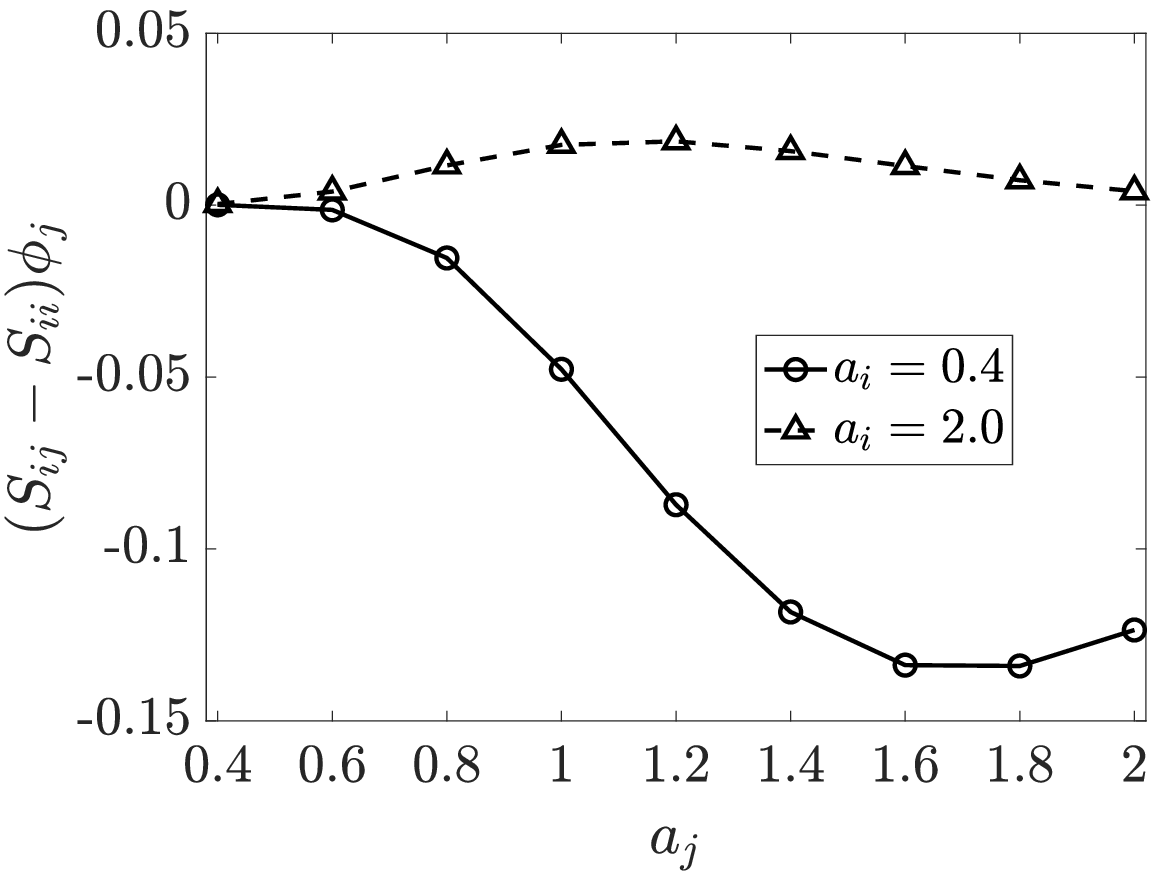}
    \caption{Inter-class interaction term appearing on the right hand side of equation (\ref{interaction}) for $a_i=0.4$ (``small'' particles) and  $a_i=2$ (``large'' particles). The particle size distribution corresponds to   $\alpha=0.4$ and $\phi=0.05$.}
    \label{fig:convolution}
\end{figure}

\begin{figure}
    \centering
    \includegraphics[width=1.0\textwidth]{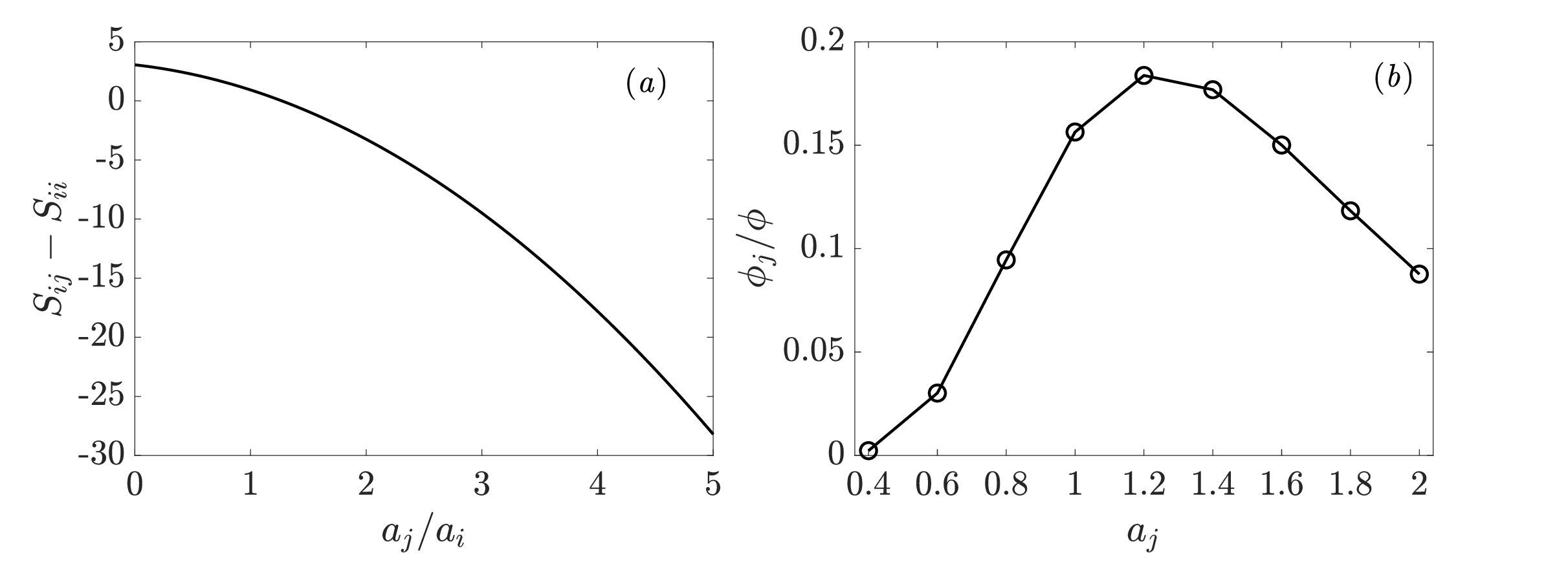}
    \caption{(\textit{a}) Interaction coefficients $S_{ij}-S_{ii}$ \protect\\ and (\textit{b}) volume fraction distribution  corresponding to the interaction term of figure \ref{fig:convolution}.}
    \label{fig:sij-sii-fre}
\end{figure}
The good comparison between Batchelor's model and the simulation data for sufficiently small $\phi$ enables us to use this analytical model to illustrate why the prediction of the velocity of the small particles is highly dependent on the full particle size distribution, while that of the large particles is not. Equation (\ref{batchelor}) can be rewritten as
\begin{equation}
h_i = 1 + S_{ii} \phi + \sum_{j=1}^m (S_{ij}-S_{ii}) \phi_j,
\label{interaction}
\end{equation}
where $S_{ii}=-6.55$ and $\phi$ is the total volume fraction. The influence of  particle class $j$ on particle class $i$ is negligible if $|(S_{ij}-S_{ii}) \phi_j| \ll |S_{ii} \phi|$. For $\phi=0.05$, the magnitude of the intra-class interaction term is $|S_{ii} \phi|=0.33$. Let us compare this value to the inter-class interaction term for $\phi=0.05$.   In figure \ref{fig:convolution}  $(S_{ij}-S_{ii})\phi_j$ is shown for $a_i = 0.4$ (small particles) and $a_i = 2$ (large particles), for $\alpha=0.4$. The maximum absolute value of the inter-class interaction term for the small particle is $0.13$, not negligible in comparison to $0.33$. The maximum value of the inter-class interaction term for the largest particles is instead 18 times smaller than the intra-class interaction term. The question is: why is the inter-class interaction term small for the large particles? Is this because the interaction coefficients are small in magnitude? Or because of the distribution of volume fractions? 

To reply to these questions, in figures \ref{fig:sij-sii-fre} (\textit{a}) and (\textit{b}) we show $(S_{ij}-S_{ii})$ and $\phi$ separately. It is seen that in our log-normal distribution the volume fraction corresponding to the small particles is small in comparison to that of the large particles, and tends to zero as the lower tail of the particle size distribution is approached. The quantity  $(S_{ij}-S_{ii})$ is on the other hand not diverging for $a_j/a_i \ll 1$, and is $O(1)$ in this limit. Therefore, the reason why the lower tail of the distribution has a small influence on the upper tail is that the volume fraction corresponding to the lower tail is comparatively small and is weighted by an interaction term that is not large.  This is an insight that could be applied to other particle size distributions. For example, if the particle size distribution was such that $\phi$ was comparatively large in the small particle range,  we would expect the settling velocity of the largest particles to be more affected by the smallest particles than seen in our simulations. 

 The analysis above also gives an insights into the condition for which models parameterised on the total volume fraction can be used as a first, practical approximation for the prediction of the settling of a  dilute polydisperse suspension. This approximation is reasonable when the inter-class interaction term is comparatively small. This term is small when either $\phi_j \ll 1$ for finite $S_{ij}-S_{ii}$, the case discussed above. Or when the particle size distribution is narrow so that $|S_{ij}-S_{ii}| \rightarrow 0$, the case discussed by  \citet{davis1988spreading} (see the value of $S_{ij}-S_{ii}$ for $a_j/a_i$ approaching 1 in figure \ref{fig:convolution} (\textit{a})). If deviations of $S_{ij}$ from $S_{ii}$ are small, then it can be seen from Eq. (\ref{general}) that the hindered settling function for $\phi \ll 1$ depends only on the total volume fraction.  For a dilute suspension with a narrow size distribution, the use of Richardson-Zaki's correlation is for example justified, and is not by chance that the exponent of the Richardson-Zaki correlation  ($n \simeq 5$) is numerically close to $|S_{ii}|=6.5$. This ``lucky coincidence'' was also noted by \citet{davis1988spreading}.

The comparison with the  simulation data shows that the MLB model tends to overestimate the hindered settling functions of particles with sizes smaller than the mean.  The MLB model gives  more accurate predictions than  Richardson-Zaki's correlation for comparatively large particles. MLB is also based on Richardson-Zaki's formula, but in the MLB model the formula is used to estimate the particle-fluid slip velocity, not the absolute settling velocity. 
For applications where the focus is predicting the sedimentation of the larger particles (e.g., separation of large particles from a polydisperse mixture), using the MLB model could be sufficient.  For applications where the stratification in different layers needs to be predicted (e.g. in sedimentology \citep{dorrell2010sedimentation}), using the MLB model will overestimate the fraction of the smaller particles in the sediment region. In particle size fractionation by centrifugation or sedimentation \citep{bonaccorso2013sorting,backes2016production}), using the MLB model  could give a wrong prediction of the region where most of the small particles are located, jeopardising the size fractionation protocol.

Looking at the main assumptions of the MLB model, rederived in the Appendix \ref{appA}, we can see that the model rests on two key assumptions. The first one is that the buoyancy force on each sphere in a polydisperse suspension depends on the suspension density.  The conceptual difficulty of modelling the buoyancy force on a small sphere in a suspension of spheres of a different size has been addressed in several papers \citep{gibb1991pressure,di1995hydrodynamics,ruzicka2006buoyancy,piazza2013general}. Experiments show that using the density of the suspension to evaluate the buoyancy force gives accurate predictions of the settling velocity of the test sphere when the test sphere  is comparable in size or larger than the other particles,  but can lead to inaccuracies when the test particle is smaller than the other particles \citep{poletto1995effective,rotondi2015validation}.  A second assumption in the MLB model is  that the Stokes drag correction for each size class only depends on the total volume fraction, and we have seen that this cannot be in general a good approximation.

We have not tried to improve the MLB model. However, we can give indications on possible approaches for improvement. When the MLB model was derived, no accurate drag correlations for polydisperse sphere arrays were available. Recently, numerical simulation of flow past fixed sphere arrays   \citep{van2005lattice,yin2009fluid,sarkar2009fluid,cheng2023hydrodynamic} have shown that  data for the drag on polydispersed sphere arrays can be fitted with low-order polynomials of the the first two moments of the size distribution. Improved models for poly-disperse suspensions could therefore be built starting from the MLB model but using the drag force correlation for polydispersed fixed arrays to model the slip velocity. Such models could also account for recent studies on the shear viscosity of polydisperse suspensions \citep{wagner1994viscosity,dorr2013discrete,mwasame2016modeling,pednekar2018bidisperse}, because in certain limits there is a relation between the drag on a sphere translating through a suspension and the shear viscosity of the suspension \citep{squires2010fluid}.    

For a continuum size distribution, equation (\ref{general}) which is also valid at non-negligible volume fractions, reads 
\begin{equation}
    h(a)=1+B_{self}\phi + \phi\int (B_{inter}-B_{self})v\left(a^{\prime}\right)da^{\prime},
    \label{eq:continous_version}
\end{equation}
where $v(a)=\frac{a^3 p(a)}{\int a^3 p(a) da}$ is the normalised volume fraction distribution corresponding to $p(a)$, and the coefficients $B_{self}$ and $B_{inter}$  are  averages of the mobility matrices. For a two-parameter distribution such as the log-normal, $v(a^{\prime})$ is a function of only two parameters, for example the mean $\langle a \rangle$ and the standard deviation $\Delta a$. One could therefore postulate a hindered settling function where the inter-class interaction term and the intra-class interaction terms are functions of $\phi, \langle a \rangle, \Delta a/ \langle a \rangle$. The issue is that the same functional form should ``best fit'' a wide range of particle sizes. This is a search which could benefit from the artificial intelligence methods \citep{zhang2020data,el2023logarithmic,wu2023enhancing,siddani2024investigating}. The analysis we have provided indicate some constraints on this search, for example the functional forms expected for $\phi \rightarrow 0$ or $\alpha \rightarrow 0$.

\section{Conclusions}
\label{sec:conclusion}

We quantified numerically the hindered settling function of  non-Brownian, dilute suspensions of polydisperse spheres with a log-normal size distribution, considering the effects of the polydispersity parameter $\alpha$ and the volume fraction $\phi$.

 The average settling velocity $\langle u_{z,i} \rangle$  of each particle size class is found to decrease as either $\phi$ or $\alpha$ increases. The class-averaged velocity $\langle u_{z,i} \rangle$ decays with $\phi$ or $\alpha$ faster for the smaller particles than the largest particles.  For given $\alpha$ and $\phi$, the hindered settling function $\langle u_{z,i} \rangle /u_{St,i}$ decreases as the particle size decreases, indicating that the settling of the smaller size classes is more hindered compared to that of the larger particles.

The probability distribution functions  of the horizontal and  vertical velocities of each size class tend to follow a Gaussian distribution, with the probability distributions of  horizontal velocities of different size classes collapsing onto each other for a given $\alpha$ and $\phi$. The magnitude of the horizontal and vertical velocity fluctuations for each size class increases as $\phi$ or $\alpha$ increases, and appear to follow the approximate scaling $u_i^{\prime} \sim u_{St,i} (a_i/\langle a\rangle)^{-2}$. Our simulations for the log-normally dispersed system suggest a value of about 3.5 for the anisotropy ratio between the vertical and the horizontal velocity fluctuations.  This value is comparable to the one observed in our simulations for  monodisperse or bidisperse suspensions.

To verify the accuracy of available polydisperse hindered settling function models, we compare the predictions of these models with the simulation data. We found that the Richardson-Zaki correlation, which is often used to model polydisperse suspensions, largely overestimates the hindered settling functions of the smaller particles. For $\alpha$=0.4 and $\phi$=0.05, the value predicted by Richardson-Zaki's formula for polydisperse suspensions can be up to seven times larger than the simulated value. Batchelor’s model (equation (\ref{batchelor})) gives quite accurate predictions for all size classes when $\phi \leq 0.05$, yielding discrepancies of the settling velocities that are within 10\% of the numerical results. The Davis \& Gecol model (equation \ref{dg}) and the MLB model (equation (\ref{mlb}))  give comparable predictions. Both these models tend to overestimate the hindered settling function of the smaller particles. The discrepancy between the models and the simulation data increases as $\alpha$ or $\phi$ increases. In Sec. \ref{sec:discussion}, we use Batchelor's model to analyse the conditions under which models parameterised on the total volume fraction can produce predictions of acceptable accuracy. 

Our simulations demonstrate that the modelling of polydispersed suspension is still challenging even in the dilute limit. Practically usable models such as Richardson-Zaki or MLB  work reasonably well for large particles, but give significant errors for small particles. This is a major obstacle in the size fractionation by centrifugation, for example, where one is interested in the precise estimation of the velocity of each particle class. In these applications, the small particle fraction is often the most valuable \citep{backes2016production}. 

Our results hold in the Stokes regime. For future work, particle-resolved simulations based on the solution of the Navier-Stokes equation around each particle \citep{uhlmann2014sedimentation,fornari2016sedimentation,willen2019resolved,yao2021effects,shajahan2023inertial} could be used to evaluate the first effect of fluid inertia on our low-Reynolds number observations. In the presence of fluid or particle inertia, averaging over instantaneous random configurations cannot be applied, because in the inertial case the particle velocities depend on the history of the hydrodynamic forces. Thanks to advances in computational power, it is however now possible to simulate tens of thousands particles at finite Reynolds numbers \citep{breugem2012second,schwarzmeier2023particle}, and with appropriate time averaging it should therefore be possible to obtain smooth statistics for at least some of the parameters combinations we explore. The most interesting seem to be the extreme cases, namely small deviations from uniformity and log-normals with a large variance (such as our $\alpha = 0.4$ case). In the presence of fluid inertia, strong trapping of small particles in the wake of large particles is expected.

Experimental techniques such as X-ray radiography \citep{dulanjalee2020measuring}, magnetic resonance imaging \citep{boyce2016magnetic}, or optical experiments with fluorescent particles  \citep{snabre2009size} could be used to measure the velocity of the small particle fraction in polydisperse suspensions. Machine learning techniques such as symbolic regression \citep{zhang2020data,el2023logarithmic,wu2023enhancing} could be used in combination with particle-resolved simulations \citep{yao2022particle} to extend Batchelor’s model to higher volume fractions, or to incorporate into  MLB's model information about the moments of the particle size distribution. 

\backsection[Acknowledgements]{We thank Wim-Paul Breugem and Johan Padding for valuable discussions. The computations are carried out on the Reynolds cluster in the Process \& Energy department in TU Delft.}

\backsection[Funding]{This project has received funding from the European Research Council (ERC) under the European Union’s Horizon 2020 research and innovation program (Grant Agreement No. 715475, project FLEXNANOFLOW).}

\backsection[Declaration of interests]{The authors report no conflict of interest.}

%%%%%%%%%%%%%%%%%%%%
\appendix

%%%%%%%%%%%%%%%%%%%%%%%%%%%%%%%%%%%%%%%%%%%%%%%%%%%%%%%%%%%%%%%%%%%%%%%%%%%%%%
\section{Derivation of the slip velocity closure in the MLB model}\label{appA}

A derivation of the MLB model is provided here to highlight the key assumptions of the model, which was too concisely described in the original papers \citep{masliyah1979hindered,lockett1979sedimentation}. Consider a homogeneous polydisperse suspension with $m$ particulate classes. The radius and density of the $j$-th class are $a_j$ and $\rho_j$, respectively, with $j=1,2,…,m$. The density and dynamic viscosity of the fluid are $\rho_f$ and $\mu$, respectively. Gravity is in the negative $z$ direction. Due to the differences between the particle and the fluid densities, a macroscopic pressure gradient $dp/dz$ along the height of the mixture is needed to balance the excess weight of the particles. This pressure gradient drives the back flow of the fluid during settling of the particles. Corresponding to this pressure gradient, each particle experiences a buoyancy force $F_{\nabla p}=(-dp/dz) V_p$, where $V_p$ is the volume of that particle. The total force exerted on each particle by the fluid is given by $F_{\nabla p}$, by the buoyancy force due to the undisturbed hydrostatic pressure gradient and by  the drag force due to the relative fluid-particle velocity difference. 

The steady-state momentum equation for the fluid phase is
\begin{equation}
    \left( -\frac{dp}{dz}\right)(1-\phi)-\sum_{j=1}^m f_{d,j}=0,
    \label{fluid-mom}
\end{equation}
where $\phi$ is the total volume fraction, and $f_{d,j}$ is the volumetric drag force density (drag per unit volume) exerted  by the $j$-th particle class.   The steady-state particle momentum equation for the $j$-th particle class is
\begin{equation}
    \left( -\frac{dp}{dz}\right)\phi_j+f_{d,j}-(\rho_j-\rho_f)\phi_j g=0,
    \label{particle-mom}
\end{equation}
where $\phi_j$ is the volume fraction of the $j$-th class. Using equations (\ref{fluid-mom}) and (\ref{particle-mom}) gives
\begin{equation}
    \frac{dp}{dz}=-\sum_{j=1}^m (\rho_j-\rho_f)\phi_j g,
    \label{pressure}
\end{equation}
and
\begin{equation}
    f_{d,j}=(\rho_j-\rho_{susp})\phi_jg,
    \label{drag-j}
\end{equation}
where $\rho_{susp}=(1-\phi)\rho_f+\sum_{i=1}^m\rho_i\phi_i$ is the density of the suspension (see e.g.  \citet{xia2022drag} for the case $m=1$). The predictive accuracy of equation (\ref{drag-j}) for small particles immersed in a suspension of larger particles has been put into question \citep{poletto1995effective,rotondi2015validation}.

To calculate the particle velocity, a constitutive equation relating relative velocity to force must be postulated. The MLB model uses a linear law between the drag force and the slip velocity $u_{slip,j}$ between the $j$-th particle class and the average fluid velocity: 
\begin{equation}
    f_{d,j}=-\beta_{j}u_{slip,j}.
    \label{drag-slip}
\end{equation}
where $u_{slip,j}$ is defined as in equation (\ref{slipdefinition}). 
The friction coefficient was calculated as $\beta_j=\frac{9\mu\phi_jC(\phi)}{2a_j^2}$.
The case $C=1$ corresponds to no influence of neighbouring particles on the drag force exerted on a test particle (the factor $\phi_j$ is due to the fact that $f_{d,j}$ is a force per unit volume). To model hydrodynamic interactions on the drag force, the MLB model assumes 
$C(\phi)=(1-\phi)^{2-n}$, as for a monodisperse case at the same total volume fraction (from Richardson-Zaki's correlation, the slip velocity in the monodisperse case is $u_{slip}=\langle u_p \rangle - \langle u_f \rangle = \frac{\langle u_p \rangle}{1-\phi}=u_{St}(1-\phi)^{n-1}$; equating (\ref{drag-j}) and (\ref{drag-slip}) using this slip velocity gives $C(\phi)=(1-\phi)^{2-n}$). 

From (\ref{drag-j}) and (\ref{drag-slip}), the slip velocity for the polydispersed case is
\begin{equation}
    u_{slip,j}=\frac{2a_j^2}{9\mu}(1-\phi)^{n-2}(\rho_j-\rho_{susp}).
    \label{slip-j}
\end{equation}
 If all the particles have the same density, $\rho_{susp} = (1-\phi) \rho_f + \phi \rho_p$ and $\rho_j - \rho_{susp} = (1-\phi) (\rho_p-\rho_f)$. In this case the slip velocity simplifies to 
\begin{equation}
    u_{slip,j}=u_{St,j}(1-\phi)^{n-1},
    \label{slip-equal-density}
\end{equation}
where $u_{St,j}$ is the Stokes velocity of the $j$-th species. Using the definition of the slip velocity and using mass continuity $\sum_j \phi_j \langle u_j \rangle + (1-\phi)\langle u_f \rangle =0$ yields equation (\ref{mlb}).  

It can be seen from the derivation that the main assumptions in MLB's model are embedded in equations (\ref{drag-j}) and (\ref{drag-slip}). 

\bibliographystyle{jfm}
\bibliography{jfm}

\begin{thebibliography}{88}
\expandafter\ifx\csname natexlab\endcsname\relax\def\natexlab#1{#1}\fi
\def\au#1{#1} \def\ed#1{#1} \def\yr#1{#1}\def\at#1{#1}\def\jt#1{\textit{#1}} \def\bt#1{#1}\def\bvol#1{\textbf{#1}} \def\vol#1{#1} \def\pg#1{#1} \def\publ#1{#1}\def\arxiv#1{#1}\def\org#1{#1}\def\st#1{\textit{#1}}

\bibitem[Abbas {\em et~al.\/}(2006)Abbas, Climent, Simonin \& Maxey]{abbas2006dynamics}
{\sc \au{Abbas, M.}, \au{Climent, E.}, \au{Simonin, O.} \& \au{Maxey, M.~R.}} \yr{2006}  \at{Dynamics of bidisperse suspensions under {S}tokes flows: Linear shear flow and sedimentation}.  \jt{Phys. Fluids}  \bvol{18}~(12).

\bibitem[Abeynaike {\em et~al.\/}(2012)Abeynaike, Sederman, Khan, Johns, Davidson \& Mackley]{abeynaike2012experimental}
{\sc \au{Abeynaike, A.}, \au{Sederman, A.J.}, \au{Khan, Y.}, \au{Johns, M.L.}, \au{Davidson, J.F.} \& \au{Mackley, M.R.}} \yr{2012}  \at{The experimental measurement and modelling of sedimentation and creaming for glycerol/biodiesel droplet dispersions}.  \jt{Chem. Eng. Sci.}  \bvol{79},  \pg{125--137}.

\bibitem[Al-Naafa \& Selim(1992)]{al1992sedimentation}
{\sc \au{Al-Naafa, M.A.} \& \au{Selim, M.~S.}} \yr{1992}  \at{Sedimentation of monodisperse and bidisperse hard-sphere colloidal suspensions}.  \jt{AIChE J.}  \bvol{38}~(10),  \pg{1618--1630}.

\bibitem[Backes(2016)]{backes2016production}
{\sc \au{Backes, C. et~al.}} \yr{2016}  \at{Production of highly monolayer enriched dispersions of liquid-exfoliated nanosheets by liquid cascade centrifugation}.  \jt{ACS Nano}  \bvol{10}~(1),  \pg{1589--1601}.

\bibitem[Batchelor(1972)]{batchelor1972sedimentation}
{\sc \au{Batchelor, G.K.}} \yr{1972}  \at{Sedimentation in a dilute dispersion of spheres}.  \jt{J. Fluid Mech.}  \bvol{52}~(2),  \pg{245--268}.

\bibitem[Batchelor(1982)]{batchelor1982sedimentation}
{\sc \au{Batchelor, G.K.}} \yr{1982}  \at{Sedimentation in a dilute polydisperse system of interacting spheres. part 1. general theory}.  \jt{J. Fluid Mech.}  \bvol{119},  \pg{379--408}.

\bibitem[Batchelor \& Wen(1982)]{batchelor1982sedimentationb}
{\sc \au{Batchelor, G.K.} \& \au{Wen, C.S.}} \yr{1982}  \at{Sedimentation in a dilute polydisperse system of interacting spheres. part 2. numerical results}.  \jt{J. Fluid Mech.}  \bvol{124},  \pg{495--528}.

\bibitem[Beenakker(1986)]{beenakker1986ewald}
{\sc \au{Beenakker, C.W.J.}} \yr{1986}  \at{Ewald sum of the rotne--prager tensor}.  \jt{J. Chem. Phys.}  \bvol{85}~(3),  \pg{1581--1582}.

\bibitem[Berres {\em et~al.\/}(2005)Berres, B{\"u}rger \& Tory]{berres2005applications}
{\sc \au{Berres, S.}, \au{B{\"u}rger, R.} \& \au{Tory, E.~M.}} \yr{2005}  \at{Applications of polydisperse sedimentation models}.  \jt{Chem. Eng. J.}  \bvol{111}~(2-3),  \pg{105--117}.

\bibitem[Bonaccorso {\em et~al.\/}(2013)Bonaccorso, Zerbetto, Ferrari \& Amendola]{bonaccorso2013sorting}
{\sc \au{Bonaccorso, F.}, \au{Zerbetto, M.}, \au{Ferrari, A.~C.} \& \au{Amendola, V.}} \yr{2013}  \at{Sorting nanoparticles by centrifugal fields in clean media}.  \jt{J. Phys. Chem. C}  \bvol{117}~(25),  \pg{13217--13229}.

\bibitem[Boyce {\em et~al.\/}(2016)Boyce, Rice, Ozel, Davidson, Sederman, Gladden, Sundaresan, Dennis \& Holland]{boyce2016magnetic}
{\sc \au{Boyce, C.~M.}, \au{Rice, N.P.}, \au{Ozel, A.}, \au{Davidson, J.F.}, \au{Sederman, A.J.}, \au{Gladden, L.F.}, \au{Sundaresan, S.}, \au{Dennis, J.S.} \& \au{Holland, D.J.}} \yr{2016}  \at{Magnetic resonance characterization of coupled gas and particle dynamics in a bubbling fluidized bed}.  \jt{Phys. Rev. Fluids}  \bvol{1}~(7),  \pg{074201}.

\bibitem[Brady \& Bossis(1988)]{brady1988stokesian}
{\sc \au{Brady, J.~F.} \& \au{Bossis, G.}} \yr{1988}  \at{Stokesian dynamics}.  \jt{Annu. Rev. Fluid Mech.}  \bvol{20}~(1),  \pg{111--157}.

\bibitem[Brady \& Durlofsky(1988)]{brady1988sedimentation}
{\sc \au{Brady, J.~F.} \& \au{Durlofsky, L.~J.}} \yr{1988}  \at{The sedimentation rate of disordered suspensions}.  \jt{Phys. Fluids}  \bvol{31}~(4),  \pg{717--727}.

\bibitem[Brady {\em et~al.\/}(1988)Brady, Phillips, Lester \& Bossis]{brady1988dynamic}
{\sc \au{Brady, J.~F.}, \au{Phillips, R.~J.}, \au{Lester, J.~C.} \& \au{Bossis, G.}} \yr{1988}  \at{Dynamic simulation of hydrodynamically interacting suspensions}.  \jt{J. Fluid Mech.}  \bvol{195},  \pg{257--280}.

\bibitem[Breugem(2012)]{breugem2012second}
{\sc \au{Breugem, W.}} \yr{2012}  \at{A second-order accurate immersed boundary method for fully resolved simulations of particle-laden flows}.  \jt{J. Comput. Phys.}  \bvol{231}~(13),  \pg{4469--4498}.

\bibitem[Brzinski~III \& Durian(2018)]{brzinski2018observation}
{\sc \au{Brzinski~III, T.A.} \& \au{Durian, D.J.}} \yr{2018}  \at{Observation of two branches in the hindered settling function at low {R}eynolds number}.  \jt{Phys. Rev. Fluids}  \bvol{3}~(12),  \pg{124303}.

\bibitem[Bürger {\em et~al.\/}(2002)Bürger, Karlsen, Tory \& Wendland]{buerger2002model}
{\sc \au{Bürger, R.}, \au{Karlsen, K.~H.}, \au{Tory, E.M.} \& \au{Wendland, W.L.}} \yr{2002}  \at{Model equations and instability regions for the sedimentation of polydisperse suspensions of spheres}.  \jt{Z Angew Math Mech.}  \bvol{82}~(10),  \pg{699--722}.

\bibitem[Chaturvedi {\em et~al.\/}(2018)Chaturvedi, Ma, Brown, Zhao \& Schuck]{chaturvedi2018measuring}
{\sc \au{Chaturvedi, S.~K.}, \au{Ma, J.}, \au{Brown, P.~H.}, \au{Zhao, H.} \& \au{Schuck, P.}} \yr{2018}  \at{Measuring macromolecular size distributions and interactions at high concentrations by sedimentation velocity}.  \jt{Nat. Commun.}  \bvol{9}~(1),  \pg{4415}.

\bibitem[Chen {\em et~al.\/}(2023)Chen, Jia, Fairweather \& Hunter]{chen2023characterising}
{\sc \au{Chen, H.}, \au{Jia, X.}, \au{Fairweather, M.} \& \au{Hunter, T.~N.}} \yr{2023}  \at{Characterising the sedimentation of bidisperse colloidal silica using analytical centrifugation}.  \jt{Adv. Powder Technol.}  \bvol{34}~(2),  \pg{103950}.

\bibitem[Cheng \& Wachs(2023)]{cheng2023hydrodynamic}
{\sc \au{Cheng, Z.} \& \au{Wachs, A.}} \yr{2023}  \at{Hydrodynamic force and torque fluctuations in a random array of polydisperse stationary spheres}.  \jt{Int. J. Multiph. Flow}  \bvol{167},  \pg{104524}.

\bibitem[Cunha {\em et~al.\/}(2002)Cunha, Abade, Sousa \& Hinch]{cunha2002modeling}
{\sc \au{Cunha, F.R.}, \au{Abade, G.C.}, \au{Sousa, A.J.} \& \au{Hinch, E.J.}} \yr{2002}  \at{Modeling and direct simulation of velocity fluctuations and particle-velocity correlations in sedimentation}.  \jt{J. Fluids Eng.}  \bvol{124}~(4),  \pg{957--968}.

\bibitem[Davis \& Birdsell(1988)]{davis1988hindered}
{\sc \au{Davis, R.H.} \& \au{Birdsell, K.H.}} \yr{1988}  \at{Hindered settling of semidilute monodisperse and polydisperse suspensions}.  \jt{AIChE J.}  \bvol{34}~(1),  \pg{123--129}.

\bibitem[Davis \& Acrivos(1985)]{davis1985sedimentation}
{\sc \au{Davis, R.~H.} \& \au{Acrivos, A.}} \yr{1985}  \at{Sedimentation of noncolloidal particles at low {R}eynolds numbers}.  \jt{Annu. Rev. Fluid Mech.}  \bvol{17}~(1),  \pg{91--118}.

\bibitem[Davis \& Gecol(1994)]{davis1994hindered}
{\sc \au{Davis, R.~H.} \& \au{Gecol, H.}} \yr{1994}  \at{Hindered settling function with no empirical parameters for polydisperse suspensions}.  \jt{AIChE J.}  \bvol{40}~(3),  \pg{570--575}.

\bibitem[Davis \& Hassen(1988)]{davis1988spreading}
{\sc \au{Davis, R.~H.} \& \au{Hassen, M.~A.}} \yr{1988}  \at{Spreading of the interface at the top of a slightly polydisperse sedimenting suspension}.  \jt{J. Fluid Mech.}  \bvol{196},  \pg{107--134}.

\bibitem[Di~Felice(1995)]{di1995hydrodynamics}
{\sc \au{Di~Felice, R.}} \yr{1995}  \at{Hydrodynamics of liquid fluidisation}.  \jt{Chem. Eng. Sci.}  \bvol{50}~(8),  \pg{1213--1245}.

\bibitem[Di~Vaira {\em et~al.\/}(2022)Di~Vaira, {\L}aniewski-Wo{\l}{\l}k, Johnson, Aminossadati \& Leonardi]{di2022influence}
{\sc \au{Di~Vaira, N.~J.}, \au{{\L}aniewski-Wo{\l}{\l}k, {\L}.}, \au{Johnson, R.~L.}, \au{Aminossadati, S.~M.} \& \au{Leonardi, C.~R.}} \yr{2022}  \at{Influence of particle polydispersity on bulk migration and size segregation in channel flows}.  \jt{J. Fluid Mech.}  \bvol{939},  \pg{A30}.

\bibitem[D{\"o}rr {\em et~al.\/}(2013)D{\"o}rr, Sadiki \& Mehdizadeh]{dorr2013discrete}
{\sc \au{D{\"o}rr, A.}, \au{Sadiki, A.} \& \au{Mehdizadeh, A.}} \yr{2013}  \at{A discrete model for the apparent viscosity of polydisperse suspensions including maximum packing fraction}.  \jt{J. Rheol.}  \bvol{57}~(3),  \pg{743--765}.

\bibitem[Dorrell \& Hogg(2010)]{dorrell2010sedimentation}
{\sc \au{Dorrell, R.} \& \au{Hogg, A.~J.}} \yr{2010}  \at{Sedimentation of bidisperse suspensions}.  \jt{Int. J. Multiph. Flow}  \bvol{36}~(6),  \pg{481--490}.

\bibitem[Dulanjalee {\em et~al.\/}(2020)Dulanjalee, Guillard, Baker, Einav \& Marks]{dulanjalee2020measuring}
{\sc \au{Dulanjalee, E.}, \au{Guillard, F.}, \au{Baker, J.}, \au{Einav, I.} \& \au{Marks, B.}} \yr{2020}  \at{Measuring grain size fractions of bidisperse granular materials using x-ray radiography}.  \jt{Opt. Express}  \bvol{28}~(20),  \pg{29202--29211}.

\bibitem[El~Hasadi \& Padding(2023)]{el2023logarithmic}
{\sc \au{El~Hasadi, Y. M.~F} \& \au{Padding, J.~T.}} \yr{2023}  \at{Do logarithmic terms exist in the drag coefficient of a single sphere at high reynolds numbers?}  \jt{Chem. Eng. Sci.}  \bvol{265},  \pg{118195}.

\bibitem[Fornari {\em et~al.\/}(2016)Fornari, Picano \& Brandt]{fornari2016sedimentation}
{\sc \au{Fornari, W.}, \au{Picano, F.} \& \au{Brandt, L.}} \yr{2016}  \at{Sedimentation of finite-size spheres in quiescent and turbulent environments}.  \jt{J. Fluid Mech.}  \bvol{788},  \pg{640--669}.

\bibitem[Gibb(1991)]{gibb1991pressure}
{\sc \au{Gibb, J.}} \yr{1991}  \at{Pressure and viscous forces in an equilibrium fluidized suspension}.  \jt{Chem. Eng. Sci.}  \bvol{46}~(1),  \pg{377--379}.

\bibitem[Gonzalez {\em et~al.\/}(2021)Gonzalez, Aponte-Rivera \& Zia]{gonzalez2021impact}
{\sc \au{Gonzalez, E.}, \au{Aponte-Rivera, C.} \& \au{Zia, R.~N.}} \yr{2021}  \at{Impact of polydispersity and confinement on diffusion in hydrodynamically interacting colloidal suspensions}.  \jt{J. Fluid Mech.}  \bvol{925},  \pg{A35}.

\bibitem[Hase \& Powell(2001)]{hase2001calculation}
{\sc \au{Hase, K.~R.} \& \au{Powell, R.~L.}} \yr{2001}  \at{Calculation of the {E}wald summed far-field mobility functions for arbitrarily sized spherical particles in stokes flow}.  \jt{Phys. Fluids}  \bvol{13}~(1),  \pg{32--44}.

\bibitem[Hasimoto(1959)]{hasimoto1959periodic}
{\sc \au{Hasimoto, H.}} \yr{1959}  \at{On the periodic fundamental solutions of the {S}tokes equations and their application to viscous flow past a cubic array of spheres}.  \jt{J. Fluid Mech.}  \bvol{5}~(2),  \pg{317--328}.

\bibitem[Hayakawa \& Ichiki(1995)]{hayakawa1995statistical}
{\sc \au{Hayakawa, H.} \& \au{Ichiki, K.}} \yr{1995}  \at{Statistical theory of sedimentation of disordered suspensions}.  \jt{Phys. Rev. E}  \bvol{51}~(5),  \pg{R3815}.

\bibitem[He {\em et~al.\/}(2021)He, Wang, Zhu, Wang, Mao, Xue \& Shi]{he2021innovative}
{\sc \au{He, W.}, \au{Wang, Q.}, \au{Zhu, Y.}, \au{Wang, K.}, \au{Mao, J.}, \au{Xue, X.} \& \au{Shi, Y.}} \yr{2021}  \at{Innovative technology of municipal wastewater treatment for rapid sludge sedimentation and enhancing pollutants removal with nano-material}.  \jt{Bioresour. Technol.}  \bvol{324},  \pg{124675}.

\bibitem[Hinch(1988)]{hinch1988sedimentation}
{\sc \au{Hinch, E.J.}} \yr{1988}  \at{Sedimentation of small particles}.  \bt{In {\em Disorder and Mixing: Convection, Diffusion and Reaction in Random Materials and Processes\/}},  \pg{pp. 153--162}.  \publ{Springer}.

\bibitem[van~der Hoef {\em et~al.\/}(2005)van~der Hoef, Beetstra \& Kuipers]{van2005lattice}
{\sc \au{van~der Hoef, M.~A.}, \au{Beetstra, R.} \& \au{Kuipers, J.A.M.}} \yr{2005}  \at{Lattice-{B}oltzmann simulations of low-{R}eynolds-number flow past mono-and bidisperse arrays of spheres: results for the permeability and drag force}.  \jt{J. Fluid Mech.}  \bvol{528},  \pg{233--254}.

\bibitem[Howard {\em et~al.\/}(2022)Howard, Maxey \& Gallier]{howard2022bidisperse}
{\sc \au{Howard, A.~A.}, \au{Maxey, M.~R.} \& \au{Gallier, S.}} \yr{2022}  \at{Bidisperse suspension balance model}.  \jt{Phys. Rev. Fluids}  \bvol{7}~(12),  \pg{124301}.

\bibitem[Kim \& Karrila(2013)]{kim2013microhydrodynamics}
{\sc \au{Kim, S.} \& \au{Karrila, S.~J.}} \yr{2013} {\em Microhydrodynamics: principles and selected applications\/}.  \publ{Butterworth-Heinemann}.

\bibitem[Ladd(1996)]{ladd1996hydrodynamic}
{\sc \au{Ladd, A. J.~C.}} \yr{1996}  \at{Hydrodynamic screening in sedimenting suspensions of non-{B}rownian spheres}.  \jt{Phys. Rev. Lett.}  \bvol{76}~(8),  \pg{1392}.

\bibitem[Lavrenteva {\em et~al.\/}(2024)Lavrenteva, Smagin \& Nir]{lavrenteva2024shear}
{\sc \au{Lavrenteva, O.M.}, \au{Smagin, I.} \& \au{Nir, A.}} \yr{2024}  \at{Shear-induced particle migration in viscous suspensions with continuous size distribution}.  \jt{Phys. Rev. Fluids}  \bvol{9}~(2),  \pg{024305}.

\bibitem[Lockett \& Bassoon(1979)]{lockett1979sedimentation}
{\sc \au{Lockett, M.~J.} \& \au{Bassoon, K.S.}} \yr{1979}  \at{Sedimentation of binary particle mixtures}.  \jt{Powder Technol.}  \bvol{24}~(1),  \pg{1--7}.

\bibitem[Malbranche {\em et~al.\/}(2023)Malbranche, Chakraborty \& Morris]{malbranche2023shear}
{\sc \au{Malbranche, N.}, \au{Chakraborty, B.} \& \au{Morris, J.~F.}} \yr{2023}  \at{Shear thickening in dense bidisperse suspensions}.  \jt{J. Rheol.}  \bvol{67}~(1),  \pg{91--104}.

\bibitem[Masliyah(1979)]{masliyah1979hindered}
{\sc \au{Masliyah, J.~H.}} \yr{1979}  \at{Hindered settling in a multi-species particle system}.  \jt{Chem. Eng. Sci.}  \bvol{34}~(9),  \pg{1166--1168}.

\bibitem[Moncho-Jord{\'a} {\em et~al.\/}(2010)Moncho-Jord{\'a}, Louis \& Padding]{moncho2010effects}
{\sc \au{Moncho-Jord{\'a}, A.}, \au{Louis, A.A.} \& \au{Padding, J.T.}} \yr{2010}  \at{Effects of interparticle attractions on colloidal sedimentation}.  \jt{Phys. Rev. Lett.}  \bvol{104}~(6),  \pg{068301}.

\bibitem[Mwasame {\em et~al.\/}(2016)Mwasame, Wagner \& Beris]{mwasame2016modeling}
{\sc \au{Mwasame, P.~M.}, \au{Wagner, N.~J.} \& \au{Beris, A.~N.}} \yr{2016}  \at{Modeling the effects of polydispersity on the viscosity of noncolloidal hard sphere suspensions}.  \jt{J. Rheol.}  \bvol{60}~(2),  \pg{225--240}.

\bibitem[Nguyen \& Ladd(2005)]{nguyen2005sedimentation}
{\sc \au{Nguyen, N.} \& \au{Ladd, A. J.~C.}} \yr{2005}  \at{Sedimentation of hard-sphere suspensions at low {R}eynolds number}.  \jt{J. Fluid Mech.}  \bvol{525},  \pg{73--104}.

\bibitem[Padding \& Louis(2004)]{padding2004hydrodynamic}
{\sc \au{Padding, J.~T.} \& \au{Louis, A.~A.}} \yr{2004}  \at{Hydrodynamic and {B}rownian fluctuations in sedimenting suspensions}.  \jt{Phys. Rev. Lett.}  \bvol{93}~(22),  \pg{220601}.

\bibitem[Papuga {\em et~al.\/}(2021)Papuga, Kaszubkiewicz \& Kawa{\l}ko]{papuga2021we}
{\sc \au{Papuga, K.}, \au{Kaszubkiewicz, J.} \& \au{Kawa{\l}ko, D.}} \yr{2021}  \at{Do we have to use suspensions with low concentrations in determination of particle size distribution by sedimentation methods?}  \jt{Powder Technol.}  \bvol{389},  \pg{507--521}.

\bibitem[Pednekar {\em et~al.\/}(2018)Pednekar, Chun \& Morris]{pednekar2018bidisperse}
{\sc \au{Pednekar, S.}, \au{Chun, J.} \& \au{Morris, J.~F.}} \yr{2018}  \at{Bidisperse and polydisperse suspension rheology at large solid fraction}.  \jt{J. Rheol.}  \bvol{62}~(2),  \pg{513--526}.

\bibitem[Peysson \& Guazzelli(1999)]{peysson1999velocity}
{\sc \au{Peysson, Y.} \& \au{Guazzelli, E.}} \yr{1999}  \at{Velocity fluctuations in a bidisperse sedimenting suspension}.  \jt{Phys. Fluids}  \bvol{11}~(7),  \pg{1953--1955}.

\bibitem[Phillips {\em et~al.\/}(1988)Phillips, Brady \& Bossis]{phillips1988hydrodynamic}
{\sc \au{Phillips, R.~J.}, \au{Brady, J.~F.} \& \au{Bossis, G.}} \yr{1988}  \at{Hydrodynamic transport properties of hard-sphere dispersions. i. suspensions of freely mobile particles}.  \jt{Phys. Fluids}  \bvol{31}~(12),  \pg{3462--3472}.

\bibitem[Piazza {\em et~al.\/}(2013)Piazza, Buzzaccaro, Secchi \& Parola]{piazza2013general}
{\sc \au{Piazza, R.}, \au{Buzzaccaro, S.}, \au{Secchi, E.} \& \au{Parola, A.}} \yr{2013}  \at{On the general concept of buoyancy in sedimentation and ultracentrifugation}.  \jt{Phys. Biol.}  \bvol{10}~(4),  \pg{045005}.

\bibitem[Poletto \& Joseph(1995)]{poletto1995effective}
{\sc \au{Poletto, M.} \& \au{Joseph, D.~D.}} \yr{1995}  \at{Effective density and viscosity of a suspension}.  \jt{J. Rheol.}  \bvol{39}~(2),  \pg{323--343}.

\bibitem[Rettinger {\em et~al.\/}(2022)Rettinger, Eibl, R{\"u}de \& Vowinckel]{rettinger2022rheology}
{\sc \au{Rettinger, C.}, \au{Eibl, S.}, \au{R{\"u}de, U.} \& \au{Vowinckel, B.}} \yr{2022}  \at{Rheology of mobile sediment beds in laminar shear flow: effects of creep and polydispersity}.  \jt{J. Fluid Mech.}  \bvol{932},  \pg{A1}.

\bibitem[Revay \& Higdon(1992)]{revay1992numerical}
{\sc \au{Revay, J.~M.} \& \au{Higdon, J. J.~L.}} \yr{1992}  \at{Numerical simulation of polydisperse sedimentation: equal-sized spheres}.  \jt{J. Fluid Mech.}  \bvol{243},  \pg{15--32}.

\bibitem[Richardson \& Zaki(1954)]{richardson1954sedimentation}
{\sc \au{Richardson, J.F.~T.} \& \au{Zaki, W.N.}} \yr{1954}  \at{Sedimentation and fluidization: part 1}.  \jt{Trans. Inst. Chem. Eng.}  \bvol{32},  \pg{35--53}.

\bibitem[Rotne \& Prager(1969)]{rotne1969variational}
{\sc \au{Rotne, J.} \& \au{Prager, S.}} \yr{1969}  \at{Variational treatment of hydrodynamic interaction in polymers}.  \jt{J. Chem. Phys.}  \bvol{50}~(11),  \pg{4831--4837}.

\bibitem[Rotondi {\em et~al.\/}(2015)Rotondi, Di~Felice \& Pagliai]{rotondi2015validation}
{\sc \au{Rotondi, M.}, \au{Di~Felice, R.} \& \au{Pagliai, P.}} \yr{2015}  \at{Validation of fluid--particle interaction force relationships in binary-solid suspensions}.  \jt{Particuology}  \bvol{23},  \pg{40--48}.

\bibitem[Ruzicka(2006)]{ruzicka2006buoyancy}
{\sc \au{Ruzicka, M.~C.}} \yr{2006}  \at{On buoyancy in dispersion}.  \jt{Chem. Eng. Sci.}  \bvol{61}~(8),  \pg{2437--2446}.

\bibitem[Sarkar {\em et~al.\/}(2009)Sarkar, van~der Hoef \& Kuipers]{sarkar2009fluid}
{\sc \au{Sarkar, S.}, \au{van~der Hoef, M.~A.} \& \au{Kuipers, J. A.~M.}} \yr{2009}  \at{Fluid--particle interaction from lattice {B}oltzmann simulations for flow through polydisperse random arrays of spheres}.  \jt{Chem. Eng. Sci.}  \bvol{64}~(11),  \pg{2683--2691}.

\bibitem[Schwarzmeier {\em et~al.\/}(2023)Schwarzmeier, Rettinger, Kemmler, Plewinski, N{\'u}{\~n}ez-Gonz{\'a}lez, K{\"o}stler, R{\"u}de \& Vowinckel]{schwarzmeier2023particle}
{\sc \au{Schwarzmeier, C.}, \au{Rettinger, C.}, \au{Kemmler, S.}, \au{Plewinski, J.}, \au{N{\'u}{\~n}ez-Gonz{\'a}lez, F.}, \au{K{\"o}stler, H.}, \au{R{\"u}de, U.} \& \au{Vowinckel, B.}} \yr{2023}  \at{Particle-resolved simulation of antidunes in free-surface flows}.  \jt{J. Fluid Mech.}  \bvol{961},  \pg{R1}.

\bibitem[Segre {\em et~al.\/}(1997)Segre, Herbolzheimer \& Chaikin]{segre1997long}
{\sc \au{Segre, P.~N.}, \au{Herbolzheimer, E.} \& \au{Chaikin, P.~M.}} \yr{1997}  \at{Long-range correlations in sedimentation}.  \jt{Phys. Rev. Lett.}  \bvol{79}~(13),  \pg{2574}.

\bibitem[Shajahan \& Breugem(2023)]{shajahan2023inertial}
{\sc \au{Shajahan, T.} \& \au{Breugem, W.}} \yr{2023}  \at{Inertial effects in sedimenting suspensions of solid spheres in a liquid}.  \jt{Int. J. Multiph. Flow}  \bvol{166},  \pg{104498}.

\bibitem[Siddani {\em et~al.\/}(2024)Siddani, Balachandar, Zhou \& Subramaniam]{siddani2024investigating}
{\sc \au{Siddani, B.}, \au{Balachandar, S.}, \au{Zhou, J.} \& \au{Subramaniam, S.}} \yr{2024}  \at{Investigating the influence of particle distribution on force and torque statistics using hierarchical machine learning}.  \jt{AIChE J.}  \pg{p. e18339}.

\bibitem[Snabre {\em et~al.\/}(2009)Snabre, Pouligny, Metayer \& Nadal]{snabre2009size}
{\sc \au{Snabre, P.}, \au{Pouligny, B.}, \au{Metayer, C.} \& \au{Nadal, F.}} \yr{2009}  \at{Size segregation and particle velocity fluctuations in settling concentrated suspensions}.  \jt{Rheol. Acta}  \bvol{48},  \pg{855--870}.

\bibitem[Squires \& Mason(2010)]{squires2010fluid}
{\sc \au{Squires, T.~M.} \& \au{Mason, T.~G.}} \yr{2010}  \at{Fluid mechanics of microrheology}.  \jt{Annu. Rev. Fluid Mech.}  \bvol{42},  \pg{413--438}.

\bibitem[Tee {\em et~al.\/}(2002)Tee, Mucha, Cipelletti, Manley, Brenner, Segre \& Weitz]{tee2002nonuniversal}
{\sc \au{Tee, S.}, \au{Mucha, P.~J.}, \au{Cipelletti, L.}, \au{Manley, S.}, \au{Brenner, M.~P.}, \au{Segre, P.~N.} \& \au{Weitz, D.~A.}} \yr{2002}  \at{Nonuniversal velocity fluctuations of sedimenting particles}.  \jt{Phys. Rev. Lett.}  \bvol{89}~(5),  \pg{054501}.

\bibitem[Tory {\em et~al.\/}(2003)Tory, Karlsen, B{\"u}rger \& Berres]{tory2003strongly}
{\sc \au{Tory, E.~M.}, \au{Karlsen, K.~H.}, \au{B{\"u}rger, R.} \& \au{Berres, S.}} \yr{2003}  \at{Strongly degenerate parabolic-hyperbolic systems modeling polydisperse sedimentation with compression}.  \jt{SIAM J. Appl. Math.}  \bvol{64}~(1),  \pg{41--80}.

\bibitem[Uhlmann \& Doychev(2014)]{uhlmann2014sedimentation}
{\sc \au{Uhlmann, M.} \& \au{Doychev, T.}} \yr{2014}  \at{Sedimentation of a dilute suspension of rigid spheres at intermediate {G}alileo numbers: the effect of clustering upon the particle motion}.  \jt{J. Fluid Mech.}  \bvol{752},  \pg{310--348}.

\bibitem[Vowinckel {\em et~al.\/}(2019)Vowinckel, Withers, Luzzatto-Fegiz \& Meiburg]{vowinckel2019settling}
{\sc \au{Vowinckel, B.}, \au{Withers, J.}, \au{Luzzatto-Fegiz, P.} \& \au{Meiburg, E.}} \yr{2019}  \at{Settling of cohesive sediment: particle-resolved simulations}.  \jt{J. Fluid Mech.}  \bvol{858},  \pg{5--44}.

\bibitem[Wacholder \& Sather(1974)]{wacholder1974hydrodynamic}
{\sc \au{Wacholder, E.} \& \au{Sather, N.~F.}} \yr{1974}  \at{The hydrodynamic interaction of two unequal spheres moving under gravity through quiescent viscous fluid}.  \jt{J. Fluid Mech.}  \bvol{65}~(3),  \pg{417--437}.

\bibitem[Wagner \& Woutersen(1994)]{wagner1994viscosity}
{\sc \au{Wagner, N.~J.} \& \au{Woutersen, A. T. J.~M.}} \yr{1994}  \at{The viscosity of bimodal and polydisperse suspensions of hard spheres in the dilute limit}.  \jt{J. Fluid Mech.}  \bvol{278},  \pg{267--287}.

\bibitem[Wang \& Brady(2015)]{wang2015short}
{\sc \au{Wang, M.} \& \au{Brady, J.~F.}} \yr{2015}  \at{Short-time transport properties of bidisperse suspensions and porous media: A {S}tokesian dynamics study}.  \jt{J. Chem. Phys.}  \bvol{142}~(9).

\bibitem[Wang {\em et~al.\/}(2020)Wang, Wang, Liu, Zhao, Wang, Wu \& Liao]{wang2020improving}
{\sc \au{Wang, X.}, \au{Wang, J.}, \au{Liu, H.}, \au{Zhao, L.}, \au{Wang, Y.}, \au{Wu, X.} \& \au{Liao, X.}} \yr{2020}  \at{Improving the production efficiency of sweet potato starch using a newly designed sedimentation tank during starch sedimentation process}.  \jt{J. Food Process. Preserv.}  \bvol{44}~(10),  \pg{e14811}.

\bibitem[Willen \& Prosperetti(2019)]{willen2019resolved}
{\sc \au{Willen, D.~P.} \& \au{Prosperetti, A.}} \yr{2019}  \at{Resolved simulations of sedimenting suspensions of spheres}.  \jt{Phys. Rev. Fluids}  \bvol{4}~(1),  \pg{014304}.

\bibitem[Wolf(2021)]{wolf2021centrifugation}
{\sc \au{Wolf, A. et~al.}} \yr{2021}  \at{Centrifugation based separation of lithium iron phosphate ({LFP}) and carbon black for lithium-ion battery recycling}.  \jt{Chem. Eng. Process.}  \bvol{160},  \pg{108310}.

\bibitem[Wu \& Zhang(2023)]{wu2023enhancing}
{\sc \au{Wu, C.} \& \au{Zhang, Y.}} \yr{2023}  \at{Enhancing the shear-stress-transport turbulence model with symbolic regression: A generalizable and interpretable data-driven approach}.  \jt{Phys. Rev. Fluids}  \bvol{8}~(8),  \pg{084604}.

\bibitem[Xia {\em et~al.\/}(2022)Xia, Yu, Pan, Lin \& Guo]{xia2022drag}
{\sc \au{Xia, Y.}, \au{Yu, Z.}, \au{Pan, D.}, \au{Lin, Z.} \& \au{Guo, Y.}} \yr{2022}  \at{Drag model from interface-resolved simulations of particle sedimentation in a periodic domain and vertical turbulent channel flows}.  \jt{J. Fluid Mech.}  \bvol{944},  \pg{A25}.

\bibitem[Xue \& Sun(2003)]{xue2003modeling}
{\sc \au{Xue, B.} \& \au{Sun, Y.}} \yr{2003}  \at{Modeling of sedimentation of polydisperse spherical beads with a broad size distribution}.  \jt{Chem. Eng. Sci.}  \bvol{58}~(8),  \pg{1531--1543}.

\bibitem[Yao {\em et~al.\/}(2022)Yao, Biegert, Vowinckel, K{\"o}llner, Meiburg, Balachandar, Criddle \& Fringer]{yao2022particle}
{\sc \au{Yao, Y.}, \au{Biegert, E.}, \au{Vowinckel, B.}, \au{K{\"o}llner, T.}, \au{Meiburg, E.}, \au{Balachandar, S.}, \au{Criddle, C.~S.} \& \au{Fringer, O.~B}} \yr{2022}  \at{Particle-resolved simulations of four-way coupled, polydispersed, particle-laden flows}.  \jt{Int. J. Numer. Methods Fluids}  \bvol{94}~(11),  \pg{1810--1840}.

\bibitem[Yao {\em et~al.\/}(2021)Yao, Criddle \& Fringer]{yao2021effects}
{\sc \au{Yao, Y.}, \au{Criddle, C.~S.} \& \au{Fringer, O.~B.}} \yr{2021}  \at{The effects of particle clustering on hindered settling in high-concentration particle suspensions}.  \jt{J. Fluid Mech.}  \bvol{920},  \pg{A40}.

\bibitem[Yin \& Sundaresan(2009)]{yin2009fluid}
{\sc \au{Yin, X.} \& \au{Sundaresan, S.}} \yr{2009}  \at{Fluid-particle drag in low-{R}eynolds-number polydisperse gas--solid suspensions}.  \jt{AIChE J.}  \bvol{55}~(6),  \pg{1352--1368}.

\bibitem[Zhang \& Ma(2020)]{zhang2020data}
{\sc \au{Zhang, J.} \& \au{Ma, W.}} \yr{2020}  \at{Data-driven discovery of governing equations for fluid dynamics based on molecular simulation}.  \jt{J. Fluid Mech.}  \bvol{892},  \pg{A5}.

\bibitem[Zuk {\em et~al.\/}(2014)Zuk, Wajnryb, Mizerski \& Szymczak]{zuk2014rotne}
{\sc \au{Zuk, P.~J.}, \au{Wajnryb, E.}, \au{Mizerski, K.~A.} \& \au{Szymczak, P.}} \yr{2014}  \at{Rotne--{P}rager--{Y}amakawa approximation for different-sized particles in application to macromolecular bead models}.  \jt{J. Fluid Mech.}  \bvol{741},  \pg{R5}.

\end{thebibliography}

\end{document}